\documentclass[fleqn,usenatbib]{mnras}

\usepackage{newtxtext,newtxmath}

\usepackage[T1]{fontenc}
\usepackage{ae,aecompl}

\usepackage{graphicx}	
\usepackage{amsmath}	
\usepackage[dvipsnames]{xcolor}
\usepackage{diagbox}
\usepackage{xspace}   

\newcommand{\total}{669\,570\xspace}
\newcommand{\stars}{581\,149\xspace}
\newcommand{\detection}{228\,613\xspace}
\newcommand{\upperlim}{352\,536\xspace}
\newcommand{\breidablik}{\textsc{breidablik}\xspace}

\newcommand{\feh}{[Fe/H]\xspace}
\newcommand{\teff}{$\rm{T}_{\rm{eff}}$\xspace}
\newcommand{\logg}{log(g)\xspace}
\newcommand{\vr}{$\rm{v}_{\rm{rad}}$\xspace}
\newcommand{\ali}{A(Li)\xspace}
\newcommand{\kms}{km\,s$^{-1}$\xspace}
\newcommand{\eteff}{$\sigma_{\rm{T}_{\rm{eff}}}$\xspace}
\newcommand{\eali}{$\sigma_{\rm{A(Li)}}$\xspace}
\newcommand{\eew}{$\sigma_{\rm{EW}}$\xspace}

\newcommand{\edit}{}

\title[3D NLTE Li abundances in GALAH DR3]{3D NLTE Lithium abundances for late-type stars in GALAH DR3}

\author[E.~Wang]{Ella~Xi~Wang$^{1, 2}$\thanks{Email: ellawang@mso.anu.edu.au}, 
Thomas~Nordlander$^{1, 2}$, Sven~Buder$^{1, 2}$, Ioana~Ciuc\u{a}$^{1, 2, 3}$, \newauthor Alexander~Soen$^{3, 4}$,
Sarah~Martell$^{2, 5}$, Melissa~Ness$^{6, 7}$, Karin~Lind$^{8}$, \newauthor Madeleine~McKenzie$^{1, 2}$, Dennis~Stello$^{2, 5, 9, 10}$
\\
$^1$Research School of Astronomy and Astrophysics, Australian National University, Canberra, ACT 2611, Australia \\
$^2$ARC Centre of Excellence for All Sky Astrophysics in 3 Dimensions (ASTRO 3D), Australia \\
$^3$School of Computing, Australian National University, Canberra, ACT 2601, Australia \\
$^4$RIKEN Center for Advanced Intelligence Project, Tokyo 103-0027, Japan \\
$^5$School of Physics, University of New South Wales, Sydney, NSW 2052, Australia \\
$^6$Department of Astronomy, Columbia University, Pupin Physics Laboratories, New York, NY 10027, USA \\
$^7$Center for Computational Astrophysics, Flatiron Institute, 162 Fifth Avenue, New York, NY 10010, USA \\
$^8$Department of Astronomy, Stockholm University, AlbaNova University Center, SE--106 91 Stockholm, Sweden \\
$^9$Sydney Institute for Astronomy (SIfA), School of Physics, University of Sydney, Sydney, NSW 2006, Australia \\
$^{10}$Stellar Astrophysics Centre, Department of Physics and Astronomy, Aarhus University, DK-8000 Aarhus C, Denmark
}

\pubyear{2023}

\begin{document}
\label{firstpage}
\pagerange{\pageref{firstpage}--\pageref{lastpage}}
\maketitle

\begin{abstract}
Lithium’s susceptibility to burning in stellar interiors makes it an invaluable tracer for delineating the evolutionary pathways of stars, offering insights into the processes governing their development. Observationally, the complex Li production and depletion mechanisms in stars manifest themselves as Li plateaus, and as Li-enhanced and Li-depleted regions of the HR diagram. The Li-dip represents a narrow range in effective temperature close to the main-sequence turn-off, where stars have slightly super-solar masses and strongly depleted Li. To study the modification of Li through stellar evolution, we measure 3D non-local thermodynamic equilibrium (NLTE) Li abundance for \stars stars released in GALAH DR3. We describe a novel method that fits the observed spectra using a combination of 3D NLTE Li line profiles with blending metal line strength that are optimized on a star-by-star basis. Furthermore, realistic errors are determined by a Monte Carlo nested sampling algorithm which samples the posterior distribution of the fitted spectral parameters. The method is validated by recovering parameters from a synthetic spectrum and comparing to 26 stars in the Hypatia catalogue. We find \detection Li detections, and \upperlim Li upper limits. Our abundance measurements are generally lower than GALAH DR3, with a mean difference of 0.23 dex. For the first time, we trace the evolution of Li-dip stars beyond the main sequence turn-off and up the subgiant branch. This is the first 3D NLTE analysis of Li applied to a large spectroscopic survey, and opens up a new era of precision analysis of abundances for large surveys.
\end{abstract}




\section{Introduction}
The chemical history of the Milky Way is encapsulated by its stars, where FGK-type stars act as fossils due to their long lifetime and convective surface which, to first approximation, retains the abundances they were born with \citep{jofre19}. Lithium is the heaviest element formed in Big Bang nucleosynthesis (BBN); therefore, stellar Li abundances (\ali\footnote{We use the customary abundance notation where $\rm{A(Li)} \equiv \log(N_\mathrm{X} / N_\textrm H) + 12$ and $[\mathrm{X}/\mathrm{Y}] \equiv (\textrm A(\mathrm{X})-\textrm A(\mathrm{Y})) - (\textrm A(\mathrm{X})-\textrm A(\mathrm{Y}))_\odot$, with N$_\mathrm{X}$ representing the number density of element ``X''.}) can be used to determine the amount of Li initially produced in BBN \citep{cyburt16}. Old, main-sequence turn-off (MSTO) dwarf stars have been found to exhibit the same Li abundance over a range of metallicities, which is a phenomenon known as the Spite plateau \citep{spite82}. The Spite plateau abundance of Li, at \ali$\approx$2.2\,dex \citep{bonifacio97, ryan99, asplund06, melendez10, sbordone10} is a factor of 3 lower than the 2.75\,dex predicted by BBN \citep{pitrou18}; this inconsistency is referred to as the cosmological Li problem \citep{fields11}. This difference is likely due to gravitational settling and turbulent mixing in stars depleting and depositing Li below the convective surface \citep{richard05, korn07}. Through the cosmological Li problem, Li links BBN and stellar evolution.

Lithium is used to probe stellar evolution because it is fragile, burning at temperatures above 2.5\,MK \citep{basri96, chabrier96, bildsten97, ushomirsky98}. Li depletes when a star undergoes the first dredge-up due to the deepening of the convective envelope evolving from the main sequence onto the red giant branch (RGB). Li is further sharply depleted at the RGB bump \citep{lind09bump}. Due to these two depletion events, metal-poor giants also form a plateau \citep{mucciarelli12, mucciarelli14, mucciarelli22}, similar to the Spite plateau. The Li-dip is a depletion of Li for MSTO stars within a narrow temperature range of 6400--6850\,K \citep{boesgaard86, gao20}, first observed in the Hyades cluster \citep{wallerstein65}, later observed in other clusters \citep{balachandran95, burkhart00} and field stars \citep{randich99, chen01, lambert04, ramirez12, bensby18, aguilera-gomez18}, with the largest set of field stars observed from GALAH DR3 in \citet{gao20}. Models are able to reproduce the observed depletion of Li in the Hyades cluster through atomic diffusion, rotation induced mixing, and internal gravity waves \citep{talon98, montalban00}. To understand how these Li depletion mechanisms vary with stellar properties such as age and metallicity requires accurate Li abundance measurements for a large set of stars. 

Elemental abundances cannot be directly measured, instead they must be inferred from models through radiative transfer: the propagation of radiation through the stellar atmosphere \edit{\citep{asplund05, lind24}}. Radiative transfer can be solved under the assumption of local thermodynamic equilibrium (LTE), which is carried out using the Saha and Boltzmann equations to calculate electron level populations and opacities. Non-LTE (NLTE) line formation is the relaxation of LTE and influences the line opacity by including the radiation field in calculating level populations. This can change the atmospheric depth at which an absorption line is formed, producing a more realistic line shape compared to LTE line formation. Measured abundances therefore differ in NLTE compared to LTE depending on the element, stellar parameters, and line strength. For Li, the difference between measured LTE and NLTE abundance can be up to 0.4\,dex \citep{lind09}. Model stellar atmospheres are used in radiative transfer to simulate the medium that radiation propagates through. 1D hydrostatic model atmospheres are often used and approximate convection through mixing length theory, micro- and macroturbulence. These approximations are relaxed in the more realistic 3D hydrodynamic model atmospheres, which model convection from first principles \citep{stein98, freytag12}. Again, measured abundances differ in 3D compared to 1D depending on the element, stellar parameters, and line strength \citep{collet06, dobrovolskas13}. For Li, the difference between measured 1D and 3D abundances can be up to 0.2\,dex (estimated from data published in \citealt{breidablik}). 

Although 3D and NLTE effects are often discussed separately, in 3D NLTE radiative transfer these effects are coupled; meaning that the difference between measured abundance in 1D LTE compared to 3D NLTE is not a straightforward sum of the difference in LTE and NLTE with the difference in 1D and 3D \edit{\citep{klevas16, lind24}}. For Li, 3D and NLTE effects partially cancel in MSTO stars, resulting in corrections close to zero when compared to 1D LTE abundances. However, these effects do not cancel in giant stars, resulting in up to 0.5\,dex difference in measured Li abundance \citep{breidablik}. 3D NLTE models provide a Li line that matches observed Li lines more closely in shape compared to 1D LTE Li profiles, but were unfeasible to utilise outside of small samples of stars due to the computational cost associated with such models \citep{lind13, mott17, wang22}. However, 3D NLTE models are now becoming more accessible with large grids being produced and published \citep{ludwig09, magic13}. Due to the simplicity of the Li atom (see \citealt{breidablik} and the references therein), there are two such grids currently available for Li \citep{harutyunyan18, breidablik}. Through these grids, it is now possible to apply 3D NLTE measurements for Li to large spectroscopic surveys. 

\edit{Existing large-scale catalogues of Li in the literature include AMBRE, GALAH DR2, LAMOST, and Gaia-ESO. 
AMBRE measures 1D LTE Li abundances with NLTE corrections from \citet{lind09} for $\sim$7000 stars from FEROS, HARPS, and UVES. These spectra were degraded to the lowest resolution of the three spectrographs, R $=40$ 000 \citep{guiglion16}. 
GALAH DR2 measures 1D NLTE Li abundances using spectra from HERMES at R $=28$ 000 for $\sim$340 000 stars using the data-driven Cannon approach \citep{buder18}.
LAMOST measures 1D LTE Li abundances for $\sim$160 000 stars at R $\approx 7500$. 
Gaia-ESO measures 1D LTE Li abundances derived from curves of growth using spectra from UVES (R $=47$ 000) and GIRAFFE (R $\approx 17$ 000) for $\sim$38 000 stars \citep{franciosini22}.
All existing surveys explore a similar parameter space of FGKM-type stars, with most surveys stopping near \feh $=-2.5$, whereas AMBRE goes lower to \feh$=-5$. Notably, none of the existing Li catalogues are in 3D NLTE.}

GALactic Archaeology with HERMES (\citealp[GALAH;][]{desilva15}) is a large spectroscopic survey conducted with the HERMES spectrograph \citep{sheinis15} using the 2dF fibre positioning system \citep{lewis02} at the 3.9-metre Anglo-Australian Telescope. Data release 3 (DR3) provided stellar parameters and abundances of 30 elements for \total spectra \citep{DR3}. These parameters were derived through Spectroscopy Made Easy (\textsc{SME}; \citealt{valenti96, piskunov17}) with 1D \textsc{marcs} model atmospheres \citep{gustafsson08}, using non-local thermodynamic equilibrium (NLTE) radiative transfer for 12 elements \citep{amarsi20}, including Li.
However, this analysis does not consider 3D effects such as the impact of the radiation-hydrodynamics interaction on the atmospheric structure; nor does it consider the interaction between 3D and NLTE effects which are non-linear.

In this paper, we present 3D NLTE Li abundances based on GALAH DR3 spectra and stellar parameters, derived using 3D hydrodynamic model atmospheres under NLTE radiative transfer, with improved treatment of blends and error estimates. In Section~\ref{obs}, we discuss the GALAH observations and parameters used in this work. In Section~\ref{analysis}, we present the analysis method used to derive 3D NLTE Li abundances and errors. In Section ~\ref{results}, we present our measured 3D NLTE Li abundances. In Section~\ref{discussion} we compare our new abundances to the GALAH DR3 abundances. Lastly, in Section~\ref{conclusion} we summarise our findings. With \total spectra containing \stars unique stars, this is the largest set of 3D NLTE Li abundances published to date. 

\section{Observations and stellar parameters}
\label{obs}
We measure Li abundances from spectra observed for GALAH DR3, using 3D NLTE spectrum synthesis. GALAH has 4 CCDs each with discrete wavelength channels, covering 4713--4903\,\AA, 5648--5873\,\AA, 6478--6737\,\AA, and 7585--7887\,\AA, respectively \citep{desilva15}. We only consider the spectra on CCD 3 as this is where the Li 6707.814\,\AA\ line of interest falls. The resolution in CCD 3 varies with wavelength and from fibre to fibre over the range $R = 22\,000$--32\,000, and is typically $\sim$25\,500 in the Li region \citep{kos17}. 

The spectra were reduced, processed, and analysed as described by \citet{kos17} and \citet{DR3}. 
We retain the stellar parameters (\teff, \logg, and \feh) from GALAH DR3 for this work. \edit{GALAH DR3 derived 1D NLTE \teff and \feh based on SME synthetic spectra and measured \logg from Gaia DR2 \citep{lindegren18} and \textit{Hipparcos} \citep{vanleeuween07} parallaxes (see \citealt{DR3} for details). They compared their stellar parameters to accurate photometric estimates and found very good agreement for cool stars, while temperatures were underestimated by upwards of 125\,K for the warmest stars. Although the \teff values in GALAH DR3 are not based on the most accurate 3D NLTE Balmer line analyses \citep[see][]{amarsi18}, they appear to be close to the fundamental scale aside from issues with the warmest stars.} 

We use S/N adopted from \texttt{snr\_c3\_iraf}, representing the S/N per pixel measured in CCD3 which contains the Li line. In addition to these parameters, we also adopt the GALAH DR3 flags associated with these parameters. In total we analysed \total spectra, reporting on \stars stars, using stacked spectra where repeated observations of the same star were taken. 

\section{Analysis}
\label{analysis}
We derive the Li abundance based on direct synthetic line profile fits in a three step process:
\begin{enumerate}
    \item Each spectrum is fitted in a two step process, over a broad region then a narrow region. The narrow Li region is fit with a 3D NLTE \breidablik \citep{breidablik} line profile for Li, and Gaussian line profiles that represent several blending lines. 
    \item From this fit, we derive the equivalent width (EW) of the best fitting Li line, along with lower and upper errors using a Monte Carlo nested sampling algorithm. 
    \item We then evaluate the Li abundance, \ali, at this EW using \breidablik and stellar parameters from GALAH DR3 \citep{DR3}. 
\end{enumerate}
\breidablik is a set of interpolation routines made available for a grid of Li line profiles calculated in 3D NLTE that were published in \citet{breidablik}. We update \breidablik for this work as discussed in Appendix~\ref{app:b}. The following subsections discuss these steps in more detail, special cases where we deviate from this process, and other technical details of the analysis.

\edit{We have chosen a traditional spectrum fitting approach for this work as opposed to label transfer: transferring Li abundances from an existing catalogue to GALAH DR3 spectra. Whilst label transfer through the use of neural networks has been fast and accurate for spectroscopic surveys on Li \citep{nepal23}, there is not a large enough training set of 3D NLTE Li abundances with blending lines to create a sufficiently accurate model for this task. However, given a large enough crossmatch, the dataset produced in this paper can become a new training set for a label transfer model of 3D NLTE Li abundances. This opens up a new method of measuring 3D NLTE Li abundances for future surveys like WEAVE \citep{dalton16} and 4MOST \citep{dejong19}.}

\subsection{Spectrum fit}
\subsubsection{Model spectra}
The Li line is heavily blended for solar metallicity stars, which are the majority of stars in GALAH. For each spectrum, we fit a 3D NLTE Li line profile and Gaussian absorption lines to blending lines in the region. The overall model spectrum is then produced by multiplying these theoretical spectra together.

We use a 3D NLTE profile for Li because it is non-Gaussian in shape when saturated. These profiles are from \breidablik, which contains a grid of 3D NLTE L think i line profiles interpolated to arbitrary stellar parameters through radial basis functions. Each blending line is modelled by one Gaussian, given by: 
\begin{equation}
    \rm{flux} = 1-\rm{A} \exp{\left(\frac{(\lambda - \lambda_0 (1+\rm{v}_{\rm{rad}}/c))^2}{FWHM^2/4\ln(2)}\right)},
    \label{gauss}
\end{equation}
where $\rm{A}$ is the amplitude of the line, FWHM is the full width half max of the line, \vr is the radial velocity, $\lambda_0$ is the line center, and $c$ is the speed of light. The \breidablik Li line profiles are broadened using a Gaussian convolution with their own separate width (FWHM$_{\rm{Li}}$). This is because \breidablik line profiles are the result of a detailed radiative transfer solution through a 3D hydrodynamic model atmosphere, and thus have a significant intrinsic width due to stellar atmospheric effects and quantum mechanics (including thermal, convective, and microscopic collisional broadening), whilst Gaussian absorption lines have an intrinsic width of zero. 

The line lists for the blending lines were compiled from \citet{melendez12} and the VALD3 database \citep{vald3}, supplemented with newer molecular data; the wavelengths and original sources are provided in Table~\ref{tab:lines}. There is a V line at 6708.094\,\AA\ and Ce line at 6708.099\,\AA; these lines have very similar strengths and are too close to be resolved, so are modelled as one line. The Fe I 6713.742\,\AA\ line is two Fe I lines merged together. There is a prominent spectral feature mainly in giant stars at $\sim$6709\,\AA\ that we could not identify the species for. We fit 2 Gaussians to this feature, labelled as ``?1'' and ``?2'' in Table~\ref{tab:lines}.

\begin{table*}
    \centering
    \caption{The wavelengths of lines used for fitting. The Li line center is taken as the weighted sum of Li components. V and Ce are at very similar strengths and wavelengths and are therefore modelled as one line.}
    \begin{tabular}{cccc}
    \hline 
    \multicolumn{2}{c}{Broad region} & \multicolumn{2}{c}{Narrow region} \\
    Center & Reference & Center & Reference \\
    \hline
    Al I 6696.085 & \citet{eriksson63} & CN 6706.730 & \citet{mandell04} \\   
    Al I 6698.673 & \citet{eriksson63} & Fe I 6707.433 & \citet{nave94} \\ 
    Fe I 6703.565 & \citet{nave94} & CN 6707.545 & \citet{mandell04} \\ 
    Fe I 6705.101 & \citet{nave94} & Li I 6707.814 & \citet{smith98} \\
    Fe I 6710.317 & \citet{nave94} & V/Ce 6708.096 & \citet{meggers75}/\citet{palmeri00} \\
    Fe I 6711.819 & \citet{nave94} & ?1 6708.810 & This work \\ 
    Fe I 6713.095 & \citet{nave94} & ?2 6709.011 & This work \\
    Fe I 6713.742 & \citet{nave94} \\ 
    Ca I 6717.681 & \citet{risberg68} \\
    \hline
    \end{tabular}
    \label{tab:lines}
\end{table*}

\subsubsection{Spectrum fit}
\begin{figure*}
    \centering
    \includegraphics[width=\textwidth]{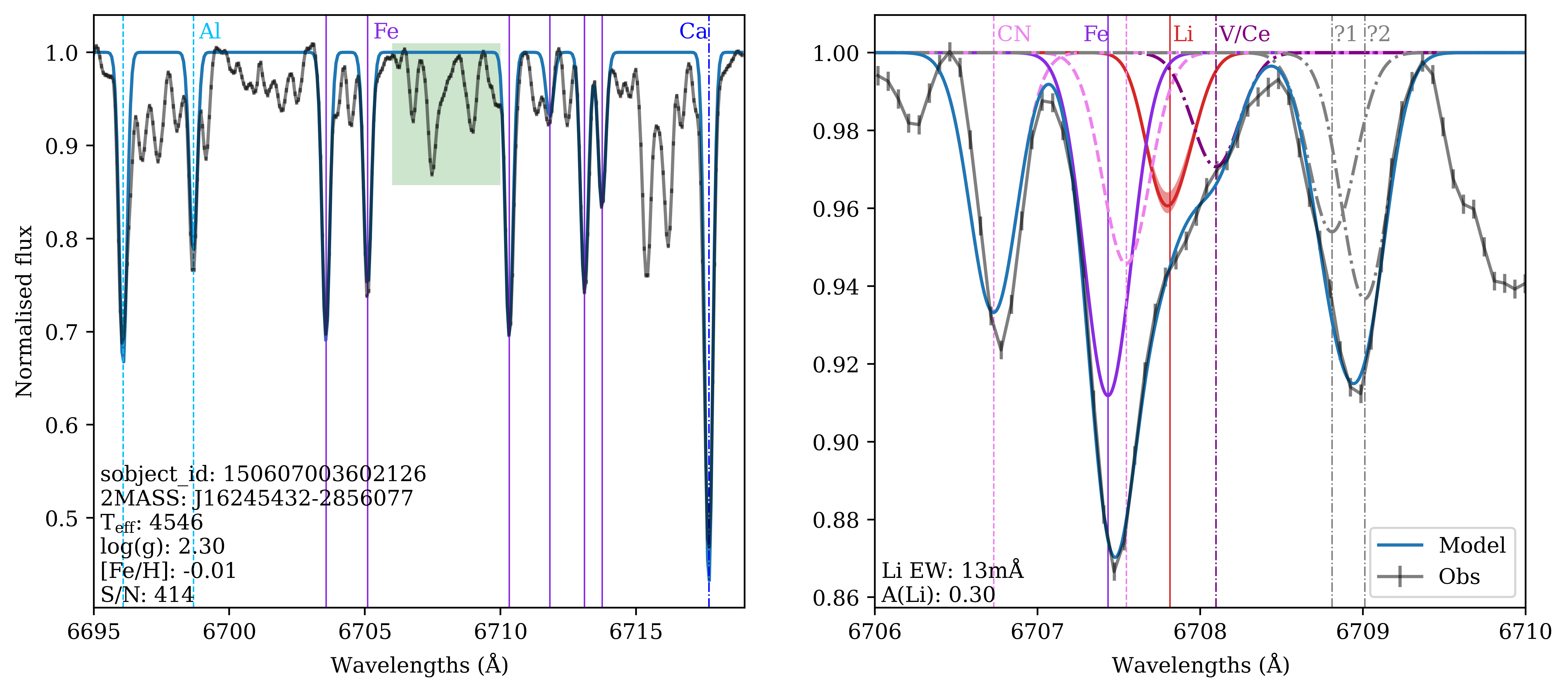}
    \caption{Fit to an example spectrum of a solar metallicity giant. Fitting of both the broad 6695--6719\,\AA\ region (left) and the narrow 6706--6710\,\AA\ region (right) corresponding to the green shaded region in the left panel is shown. Each theoretical spectrum is coloured per element, with the vertical lines showing the central wavelength of the line. The red shaded region in the right panel is the range that the synthetic spectrum would change by if there was a 1$\sigma$ change in EW. The combined theoretical spectrum is shown in blue. Spectral properties in the bottom left are from GALAH DR3.}
    \label{fig:fit}
\end{figure*}

Each spectrum is fitted in two steps using a broad (6695--6719\,\AA) and narrow (6706--6710\,\AA) region. The fits to both regions are shown in Fig.~\ref{fig:fit}. 

First, we fit the broad region, using a single value for the blending line width FWHM and radial velocity \vr based on nine Al, Fe, and Ca lines in the broad region, omitting the narrow region, shown in the left panel of Fig.~\ref{fig:fit}. Then, we fit the narrow region, where the fitted FWHM and \vr from the broad region are applied, and we now treat the continuum normalisation as a variable offset. We model the blending lines with six Gaussians, whilst Li is modelled with \breidablik line profiles broadened by FWHM$_{\rm{Li}}$, shown in the right panel of Fig.~\ref{fig:fit}. We adopt this two step approach to constrain the \vr and FWHM free from influence from the Li line, thus reducing the number of free parameters fitted over the narrow region. The 6706.730 CN line is not modelled perfectly by a Gaussian; however, because it overlaps so little with the Li line, this will not affect our derived Li EW significantly. Additional example fits of a warm solar metallicity dwarf with low S/N, a poorly constrained star, and a saturated star are shown in Appendix~\ref{app:fits}.

We fit a single value of FWHM and \vr for all lines in the spectrum. The radial velocity \vr represents the residual velocity of the Li line after the spectrum has been shifted to the rest frame, primarily due to a velocity drift caused by errors in the wavelength calibration. The width of the line, FWHM, is assumed to be the same for all lines because with the typical 10\,\kms resolution of the HERMES instrument, the line shape is dominated by the instrumental profile together with rotational broadening rather than broadening intrinsic to the line or due to convection. The fitted FWHM and \vr from this work are slightly different from the values reported in GALAH DR3 as we are only fitting these values over the 6695--6719 \AA\ region as opposed to GALAH which fits over all CCDs; and we define these values differently. In particular, FWHM from this work represents the combined rotational and instrumental broadening under the assumption of a Gaussian line profile.

To prevent unconstrained fits in stars with no measurable Li but a depressed continuum, we apply constraints on both FWHM and \vr. We take the rotational broadening value measured from GALAH DR3, and assume that the instrumental resolution lies in the range R$=$22000--32000 \citep{kos17}; and sum the rotational and instrumental broadening in quadrature as limits for FWHM. A similar upper limit is placed for FWHM$_{\rm{Li}}$, but due to the intrinsic broadening of \breidablik line profiles, the lower limit is 0. We limit \vr to prevent it from fitting noise in the continuum when there are no lines present. This limit is empirically set to half the upper FWHM limit because it is harder to measure \vr in stars with higher rotation due to increased blending.

\subsubsection{Special cases}
We deviate from the standard aforementioned fitting method for poorly constrained stars and stars outside of the \breidablik grid (see Section~\ref{sec:grid} for details on the \breidablik grid). 

A spectrum is classified as poorly constrained if fewer than three lines can be measured in the broad region. This most often occurs when stars are metal-poor, hot, or rapidly rotating. 
The three line cutoff is determined empirically. 
If a spectrum is poorly constrained, then the Li line is only fit with a \breidablik line profile assuming no contribution from other elements. The FWHM$_{\rm{Li}}$ and \vr are fitted simultaneously with Li EW as there is no strong constraint on these values from the broad region.

We also deviate from the standard fitting method for stars with stellar parameters outside of the \breidablik grid. These stellar parameters are too far from the \breidablik grid and so the interpolated 3D NLTE Li line profile is untrustworthy. Instead, we use a Gaussian to also model the Li line, with the same constraints as the blending lines. These Li EWs are reliable except for stars with saturated Li lines. 

\subsection{Equivalent width measurements}
The EW of the Li line is calculated via numerically integrating the best fitting \breidablik line profile. To account for measurement error introduced by uncertainties in the strengths of the blending lines, we sample the posterior distribution of the fit given by $\chi^2$ likelihood using the nested sampling Monte Carlo algorithm, MLFriends \citep{buchner16, buchner19}, from \textsc{ultranest} \citep{buchner21}. 

\textsc{ultranest} is able to sample posterior distributions and return converged chains free from typical tuning parameters such as burn-in and the number of samples. This is possible because \textsc{ultranest} samples a posterior distribution by iteratively shrinking the parameter space towards higher likelihoods and iteratively updating the evidence. In order to capture the full posterior distribution and to distinguish between detections and non-detections, we allow Li to be in emission. This is implemented as a reflection of the absorption line with the same EW. The blending lines can only be in absorption or else the solution is poorly constrained. 

\begin{figure}
    \centering
    \includegraphics[width=0.48\textwidth]{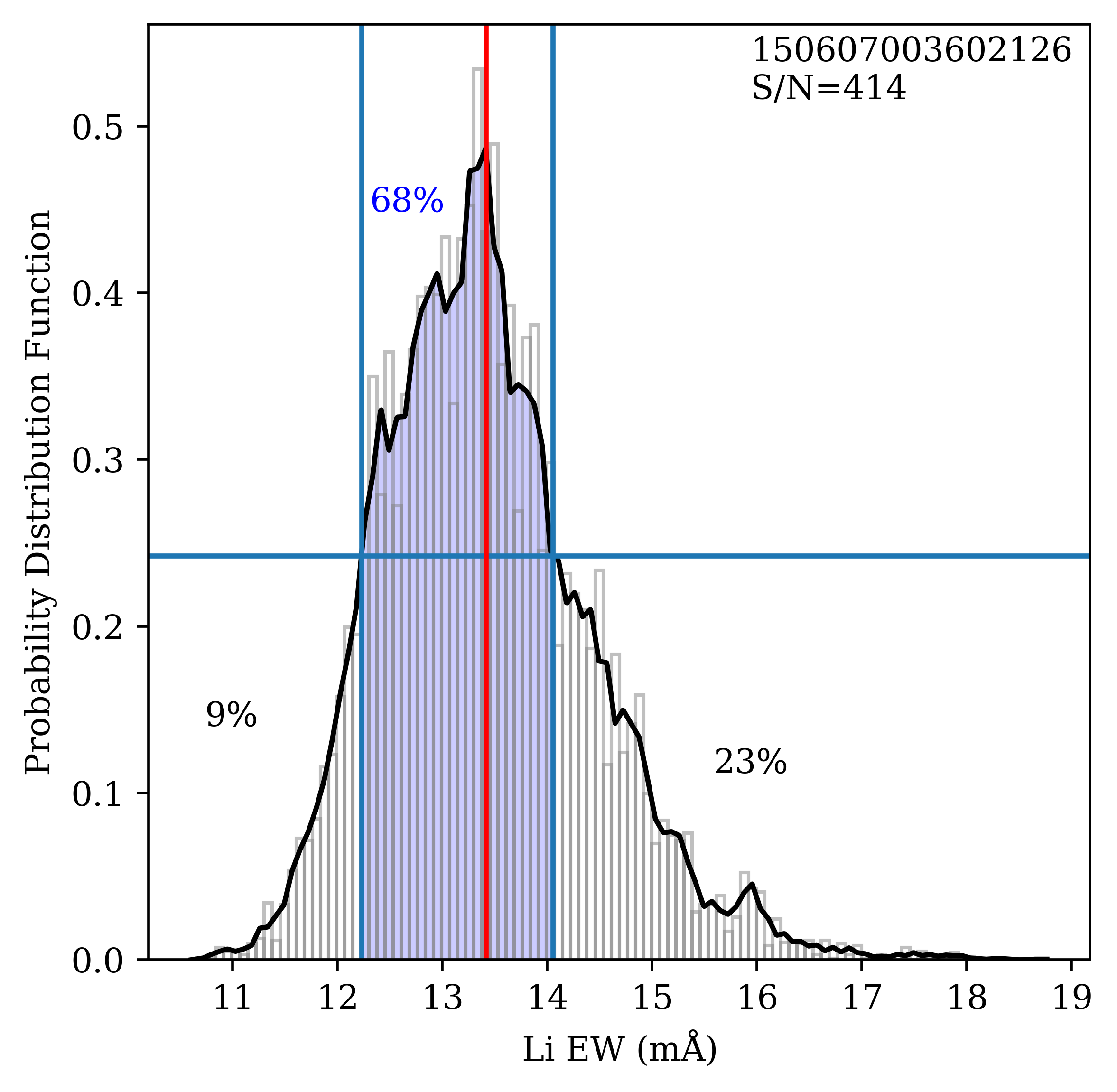}
    \caption{Example sampled posterior showing how we derive statistics. The \texttt{sobject\_id} and S/N of this star is shown in the top right corner. The histogram is the sampled posterior binned (black bars), with the smooth density (solid black line) obtained via a running mean. The measured Li EW is adopted as the MAP (red); and the lower and upper errors are adopted as the highest posterior density Bayesian credible interval (blue), where the horizontal line used to identify the error in Li EW is also shown.}
    \label{fig:err}
\end{figure}

To interpret the sampled posterior, we make a histogram using the sampled posterior, with bin widths given by the Stone algorithm \citep{stone84} because it produces reasonable bin widths for both normal and skewed distributions. We then smooth the histogram with a running mean of window size 3 to reduce noise. From this density histogram, we measure the Li EW and errors as shown in Fig.~\ref{fig:err}. The measured Li EW is adopted as the maximum a posteriori (MAP) estimate (the mode of the smoothed histogram) as this typically fits the observed spectra better than the mean; and the lower and upper errors are given by the highest posterior density Bayesian credible interval \citep{hespanhol19}. This method of analysing sampled posteriors is very similar to \citet{hinton16}.

Fig.~\ref{fig:err} shows our lower and upper errors on a well behaved distribution. 
The errors are given by the highest posterior density Bayesian credible interval: the EWs with the same probability density value that encompass a 68\% region of the sampled posterior. 
To find these EWs, first, a horizontal line is drawn, then the EWs where it intersects the sampled posterior are identified, lastly the area encompassed by these EWs are measured. This horizontal line is adjusted until the area encompassed by the EWs is 68\%, and the corresponding EWs are then the 1$\sigma$ lower and upper errors. 
The highest posterior density Bayesian credible interval reduces to a standard \( \pm 1\sigma \) confidence interval when the distribution is symmetric and unimodal. There are 170 stars where the sampled posterior is not fully captured due to restrictions from the prior, we discuss how the errors for these stars are derived in Appendix~\ref{app:err}.

\subsubsection{Sampled posterior probability distribution region}
The posterior is sampled over a bounded region. We set this bounded region using the parameters of the best fit from \texttt{scipy.optimize.minimize} adding a small error region around these parameters given by the error formula \citep{norris01}:
\begin{equation}
    \sigma_{\rm{EW}} = \lambda n^{1/2}_{\rm{pix}} / R (\textrm{S/N}),
\end{equation}
where $R$ is the resolving power, $\textrm{S/N}$ is the signal to noise ratio per pixel, and n$_{\rm{pix}}$ is the number of pixels in the line. The posterior is then well sampled if most of the probability distribution function is captured. We define this by considering where the mode of the sampled posterior falls. First, we take the histogram of the sampled posterior for one parameter using 100 bins, then we identify the mode of this histogram. If the mode is within 5 bins of either the lower or upper edge of the histogram, then this sampled posterior is not well captured for this parameter. This process is then repeated for all parameters. If the posterior is not well captured for any EWs or the continuum normalisation parameter, we sample the posterior again over a larger region. If this re-sampling still fails to capture the full posterior, then we flag this star. Flags are discussed further in Section~\ref{sec:flags}.

\subsubsection{Non-detections}
We determine whether Li is detected using a 2 sigma cut on the EW:
\begin{equation}
\rm{EW} < 2 \sigma^{\rm l}_{\rm{EW}}.
\end{equation}
For stars where the Li line is not detected, we report the abundance corresponding to \(2 \sigma^{\rm l}_{\rm{EW}} \) as the upper limit. By this definition of non-detection, negative EWs are also necessarily non-detections that we report an upper limit for. We note that errors in the model are not taken into account, so our errors are underestimated. 

In cases where $\sigma^{\rm l}_{\rm{EW}}$ does not exist, reported as NaN in the data (see right panel of Fig.~\ref{fig:err}), we the \citet{norris01} error formula to determine whether the line is detected. 
In the case of a non-detection we report upper limit 2$\sigma_{\rm{EW}}$, similar to stars with a non-NaN error, 2$\sigma^{\rm l}_{\rm{EW}}$.

\subsection{Lithium abundance measurements}
We estimate the Li abundance, \ali, from the measured EW using the 3D NLTE package \breidablik, taking the \teff, \logg, and \feh values from GALAH DR3. \breidablik interpolates from a grid of 3D NLTE Li EWs and \ali to arbitrary stellar parameters using a feedforward neural network. There are two main sources of error for \ali (\eali): error in \teff (\eteff) and error in EW (\eew). \eali due to \eteff is provided by calculating Li abundance when perturbing \teff by \eteff from GALAH DR3 and reported as the mean of the lower and upper error. \eali due to \eew is provided separately as an asymmetric error, where the lower \eali is the Li abundance corresponding to the lower error in EW ($\sigma^{\rm l}_{\rm{EW}}$), and the upper \eali is the Li abundance corresponding to the upper error in EW ($\sigma^{\rm u}_{\rm{EW}}$). \edit{We do not report the error in \ali due to error in \logg or error in \feh, as these are much smaller than the aforementioned sources of error, typically of the order of 0.01\,dex.}

\subsection{Validation}
\subsubsection{Synthetic spectra}
We validate our analysis pipeline on a randomly generated synthetic spectrum, intended to be similar to a heavily blended Solar spectrum. To create this synthetic spectrum, we synthesize a broad region spectrum and narrow region spectrum separately then multiply them together. The broad region spectrum is created through our template spectrum using EWs of $\sim$70\,m\AA. In addition, we include 40 Gaussians of EW $\sim$20\,m\AA\ centered at random wavelengths in the broad region excluding the narrow region to mimic the potential impact of blending lines that we do not model. The narrow region is then created through the template spectrum with EWs of $\sim$20\,m\AA. We add a continuum normalisation factor of 0.99 and noise corresponding to S/N$=100$. The fitted synthetic spectrum and true parameters are shown in Appendix~\ref{app:synth}.

\begin{figure*}
    \centering
    \includegraphics[width=\textwidth]{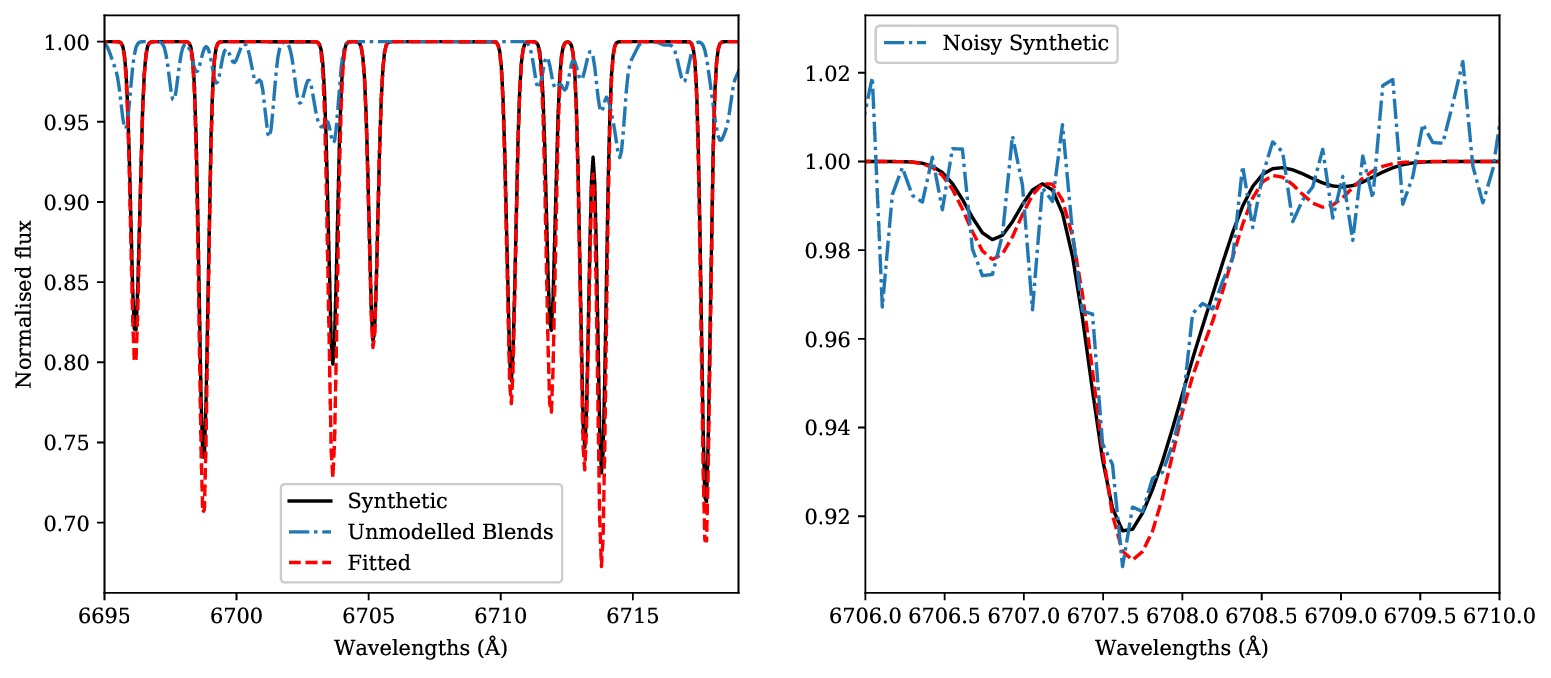}
    \caption{The components to the synthetic spectrum and the fitted spectrum. The left panel shows the broad region without noise or continuum normalisation factor, with the unmodelled blends shown. The right panel shows the narrow region without continuum normalisation factor, for both the synthetic spectrum with and without noise.}
    \label{fig:synth_break}
\end{figure*}

We show each component of the synthetic spectrum along with the fits in Fig.~\ref{fig:synth_break}. For this comparison, we remove the continuum normalisation factor from the synthetic spectrum for clarity. The left panel shows that the measured EWs are higher for all lines compared to the true EWs, and this is worse where unmodelled blends are stronger, showing unmodelled blends inflate the measured EWs. Unmodelled blends also cause the measured FWHM to be higher than the true FWHM. In addition, we do not normalise the broad region spectrum, which contributes to the overestimated parameters for this example. Whilst this does affect our Li EW measurements, it does not introduce a clear bias towards higher or lower Li EWs. Unlike the broad region, the fitted narrow region EWs can either be smaller or larger than the true EWs. instead, the fitted spectrum better follows the noisy spectrum as opposed to the synthetic spectrum, as shown in the right panel, implying that the differences in measured EW for the narrow region is due to noise rather than blending lines. We fit a Li EW of $17_{-5}^{+2}$\,m\AA, which gives \ali$=1.74$\,dex; compared to the true Li EW of 20\,m\AA\ corresponding to \ali$=1.80$\,dex. We note however that the true value almost falls within 1$\sigma$ of our measured Li EW, despite a large portion of the offset being caused by systematic rather than random errors. We recover a continuum normalisation factor of 0.995, which places the measured continuum higher than the true continuum by 0.5\%.

\subsubsection{Hypatia Catalogue}
\begin{figure}
    \centering
    \includegraphics[width=0.48\textwidth]{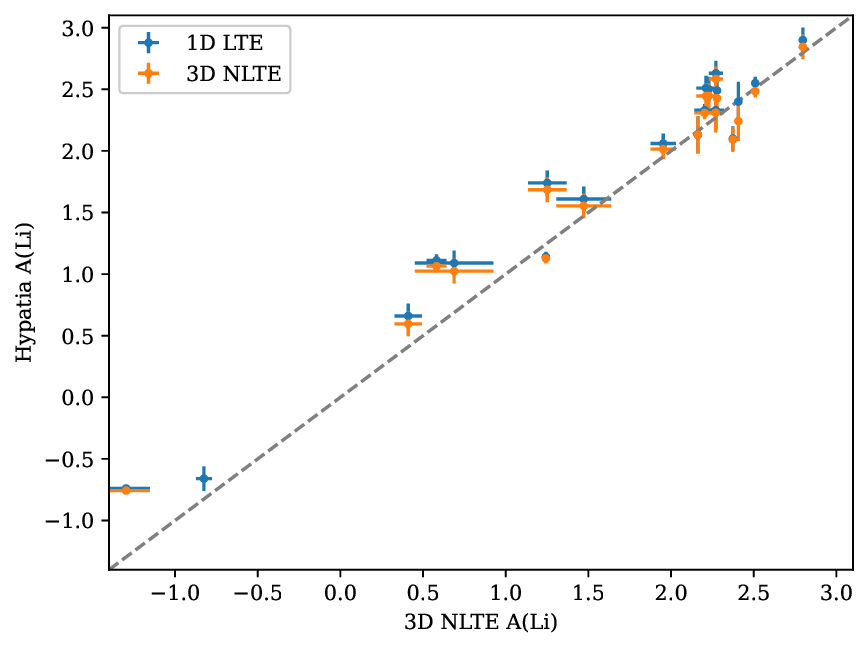}
    \caption{3D NLTE \ali from this work compared to \ali for 26 stars in common with the Hypatia \citep{hypatia} catalogue. Blue points are the 1D LTE \ali from Hypatia, orange points are these abundances in 3D NLTE, derived using the 3D NLTE corrections published in \citet{breidablik}. Our abundances are roughly in agreement with Hypatia but are generally lower even when including a 3D NLTE correction. The gray dashed line is where the abundances are equal.}
    \label{fig:hyp}
\end{figure}

We validate that our treatment of blends successfully recovers the Li line by comparing to a sample of stars from the Hypatia catalogue \citep{hypatia}, shown in Fig.~\ref{fig:hyp}. \edit{The crossmatched stars span $4280\leq$ \teff$\leq 6826$\,K, $1.31\leq$ \logg$\leq 4.58$, and $-2.42\leq$ \feh$\leq 0.57$. We crossmatch to Hypatia as this catalogue has the largest set of crossmatched stars with lithium in GALAH, but note that as the Hypatia data are collected from different sources there may be significant systematics present}. For these 26 crossmatched stars, we find similar but slightly lower abundances for 1D LTE \ali from Hypatia. we derive 3D NLTE abundance corrections from \breidablik \citep{breidablik}, and find that these remove a small amount of the offset, but is not enough to account for the full difference. We find a mean difference of 0.18\,dex, with scatter 0.21\,dex; applying 3D NLTE corrections, the mean difference becomes 0.14\,dex whilst scatter remains unchanged. These abundances are not within 1$\sigma$ agreement of each other, but it's unclear if this is because errors are underestimated on our side or in the Hypatia catalogue. We do not find any trends in abundance difference with \teff, \logg, \feh, or \ali, indicating that our results perform equally well for all stellar parameters and are not biased by the different amounts of blending present for different stars. 

\subsection{Synthetic spectra grid}
\label{sec:grid}

The synthetic spectra grids used in \breidablik and GALAH are different, and as a result, stellar parameters derived in GALAH DR3 may be outside the \breidablik grid. The GALAH DR3 analysis is based upon the 1D \textsc{marcs} grid \citep{gustafsson08} and spans $2500\leq$ \teff$\leq 8000$\,K, $-0.5\leq$ \logg$\leq 5.5$\,dex, and $-5.0\leq$ \feh$\leq 1.0$; whilst the \breidablik grid is based on the 3D \textsc{stagger} grid \citep{magic13} and spans $4000\leq$ \teff$\leq 7000$\,K, $1.5\leq$ \logg$\leq 5.0$\,dex, and $-4.0\leq$ \feh$\leq 0.5$. The Li abundance for stars outside the stellar grid are reported but flagged as untrustworthy, see Section~\ref{sec:flags} for details.

For stellar parameters near the edge of the grid, it is not obvious whether this is inside or outside the grid. We consider a star to be within the \breidablik grid if a star is within half the distance between two adjacent grid points to the nearest grid model. In practice, this is defined as the normalised stellar parameters falling within a sphere of radius $\sqrt{3 \times 0.5^2}$ of the closest regularised grid point. In 3D hydrodynamic model atmospheres, \teff is an output rather than an input, so the true \teff differ from the nominal \teff, causing the grid to be irregular in \teff. To prevent holes in the grid, we use the nominal \teff values to determine whether a star is in the grid or not. To normalise the stellar parameters, we divide the stellar parameter by their respective step sizes, 500\,K for \teff, 0.5\,dex for \logg, and 1 for \feh. 

It is possible for the measured \teff to be within the grid, and the \eteff to perturb that \teff to be outside of the grid, i.e. \teff is in the grid but one or both of $\textrm{\teff}\pm\textrm{\eteff}$ is outside of the grid. Where $\textrm{\teff}+\textrm{\eteff}$ is outside of the grid, we report \eali from $\textrm{\teff}-\textrm{\eteff}$; and vice versa. If both $\textrm{\teff}\pm\textrm{\eteff}$ fall outside of the grid, then \eali due to \eteff could not be evaluated and is reported as \texttt{NaN}. This occurs for 34 stars in our data set. 

The grid covers a range in \ali from $-0.5$\,dex to 4\,dex, however, we still report \ali and \eali outside of these limits, as extrapolation in this parameter is relatively well behaved. However, we do caution that these extrapolated abundances should be verified before use.

\subsection{Flags}
\label{sec:flags}

We use a bitmask flag (\texttt{flag\_ALi}) to indicate details with the analysis shown in Table~\ref{tab:flag_ali}, similar to GALAH DR3. For example, if a particular star is a non-detection and the Li EW posterior is not fully captured, then \texttt{flag\_ALi} will be 5. We recommend using \texttt{flag\_ALi} < 4 when utilising Li EWs, removing 219 stars, because Li EWs are well measured if the continuum is well placed and the Li EW posterior is fully captured. When utilising \ali, we recommend \texttt{flag\_ALi} < 2, removing 41939 stars outside of the \breidablik grid. This stricter flag selection is because \ali is sensitive to extrapolation in the \breidablik grid whilst Li EW is not significantly affected.

\begin{table}
    \centering
    \caption{Flags used in this analysis to calculate the final bitmask flag \texttt{flag\_ALi}.}
    \begin{tabular}{cc}
    \hline 
    \multicolumn{2}{c}{\texttt{flag\_ALi}} \\
    \hline
    1 & Non-detection, upper limit reported \\
    2 & Stellar parameters fall outside the \breidablik grid \\ 
    4 & Li EW posterior is not fully captured \\ 
    8 & Continuum posterior is not fully captured \\
    16 & Bad stellar parameter flag ($\geq$ 128) \\
    \hline
    \end{tabular}
    \label{tab:flag_ali}
\end{table}

We reproduce the GALAH flags on stellar parameters, metallicity, and \ali that are adopted in this work in Table~\ref{tab:flags}. In general, we recommend taking \texttt{flag\_sp\_DR3} and \texttt{flag\_fe\_h\_DR3} as 0, and \texttt{flag\_ALi\_DR3} $<$ 2 when using the GALAH DR3 data replicated in this work. In particular, \texttt{flag\_ALi\_DR3} $=$ 0 for detections and \texttt{flag\_ALi\_DR3} $=$ 1 for upper limits. For further details on the GALAH DR3 flags, see \citet{DR3}.

\begin{table}
    \centering
    \caption{Flags from GALAH DR3 that are adopted from \citet{DR3} in this work.}
    \begin{tabular}{cc}
    \hline 
    \multicolumn{2}{c}{\texttt{flag\_sp\_DR3}} \\
    \hline
    1 & Unreliable astrometric solution, Gaia RUWE > 1.4 \\
    2 & Unreliable broadening \\ 
    4 & Low S/N (below 10 for CCD 2) \\ 
    8 & Reduction issues \\
    16 & t-SNE projected emission features \\
    32 & t-SNE projected binaries \\
    64 & Binary sequence/pre-main sequence flag \\
    128 & S/N-dependent high \textsc{SME} $\chi^2$ (bad fit) \\
    256 & Problems with Fe \\
    512 & \textsc{SME} did not finish \\
    1024 & Stellar parameters fall outside of the \textsc{marcs} grid \\
    \hline
    \multicolumn{2}{c}{\texttt{flag\_fe\_h\_DR3}} \\
    \hline
    1 & Upper Limit \\
    2 & Bad fit / large $\chi^2$ \\
    4 & Uncertain measurement / saturation \\
    16 & Bad stellar parameter flag ($\geq$ 128) \\
    32 & No measurement available \\
    \hline
    \multicolumn{2}{c}{\texttt{flag\_ALi\_DR3}} \\
    \hline
    1 & Upper Limit \\
    2 & Bad fit / large $\chi^2$ \\
    4 & Uncertain measurement / saturation \\
    8 & Bad wavelength solution / rv for Li$_{6708}$ \\
    16 & Bad stellar parameter flag ($\geq$ 128) \\
    32 & No measurement available \\
    \hline
    \end{tabular}
    \label{tab:flags}
\end{table}

\section{Results}
\label{results}
In this work, we present a comprehensive catalogue of Li abundance measurements as a supplementary catalogue for GALAH DR3. The catalogue is tabulated as described in Table~\ref{tab:data}. \edit{In total, we used $\sim$600k CPU hours to compute this catalogue, with majority of the computational time going to posterior sampling. Due to the different number of parameters fitted for poorly constrained stars compared to the standard analysis procedure, the time per star varies between 30 minutes to 2 hours.}
We present both the ``allspec'' and ``allstar'' catalogues from GALAH DR3. ``allspec'' contains all spectra, including repeated observations of the same star; whilst ``allstar'' includes analysis from only the coadded spectra for stars that have been observed multiple times (refer to \citealt{DR3} for the treatment of duplicate spectra). All analysis of results and comparison to GALAH DR3 for this work has been done with the ``allstar'' catalogue. 

We use \teff, \logg, and \feh from GALAH DR3 to re-derive the lithium abundance. Stars that do not have stellar parameters reported in GALAH DR3 are not included in our abundance analysis. We carry over the GALAH DR3 flags and add our own flag, but we report Li EW and \ali where possible regardless of these flags.
In total, we measure \ali in 
\total spectra which contains
\stars unique stars, of which \detection are detections and \upperlim are upper limits. 

\begin{table*} 
    \centering
    \caption{Schema for the outputs of both the \texttt{allstar} and \texttt{allspec} catalogues for this work. The results from this work are made available through a Value Added Catalogue to GALAH DR3. We provide both results from this work and include relevant results from GALAH DR3.}
    \begin{tabular}{lllp{10cm}}
    \hline
    Name & Data type & Units & Description \\
    \hline
    \texttt{sobject\_id} & int & -- & Spectrum ID from GALAH DR3 \\
    \texttt{star\_id} & str & -- & Star ID from GALAH DR3 \\
    \texttt{fwhm\_broad} & float & \kms & Width of the lines measured from broad region, NaN if poorly constrained \\
    \texttt{fwhm\_Li} & float & \kms & Width of the \breidablik Li line, NaN if outside grid \\
    \texttt{vbroad\_DR3} & float & \kms & Rotational broadening from GALAH DR3 (no instrumental profile) \\
    \texttt{delta\_rv\_6708} & float & \kms & Relative \vr compared to GALAH DR3 measured from the broad region, measured from Li region if poorly constrained \\
    \texttt{rv\_DR3} & float & \kms & \vr from GALAH DR3 \\
    \texttt{EW} & float & m\AA & Measured Li EW \\
    \texttt{e\_EW\_low} & float & m\AA & Lower error in Li EW ($\sigma^{\rm l}_{\rm{EW}}$) \\ 
    \texttt{e\_EW\_upp} & float & m\AA & Upper error in Li EW ($\sigma^{\rm u}_{\rm{EW}}$) \\ 
    \texttt{e\_EW\_norris} & float & m\AA & Error using formula from \citet{norris01} \\
    \texttt{ALi} & float & dex & 3D NLTE A(Li) detections \\
    \texttt{ALi\_upp\_lim} & float & dex & 3D NLTE A(Li) upper limit corresponding to $2\sigma^{\rm l}_{\rm{EW}}$ \\
    \texttt{e\_ALi\_low} & float & dex & Lower error in 
    \ali associated with $\sigma^{\rm l}_{\rm{EW}}$ \\
    \texttt{e\_ALi\_upp} & float & dex & Upper error in \ali associated with $\sigma^{\rm u}_{\rm{EW}}$ \\
    \texttt{e\_ALi\_teff} & float & dex & Error in \ali associated with $\sigma_{\rm{T}_{\rm{eff}}}$ \\
    \texttt{flag\_ALi} & int & -- & Bitmask flag for this analysis (see Section~\ref{sec:flags}) \\
    \texttt{ALi\_DR3} & float & dex & 1D NLTE A(Li) detections and upper limits from GALAH DR3 \\
    \texttt{e\_ALi\_DR3} & float & dex & $\sigma_{\rm{A(Li)}}$ from GALAH DR3 \\
    \texttt{flag\_ALi\_DR3} & int & -- & Bitmask flag for A(Li) from GALAH DR3 \\
    \texttt{teff\_DR3} & float & K & \teff from GALAH DR3 \\
    \texttt{e\_teff\_DR3} & float & K & $\sigma_{\rm{T}_{\rm{eff}}}$ from GALAH DR3 \\
    \texttt{logg\_DR3} & float & log(cm s$^{-2}$)& \logg from GALAH DR3 \\
    \texttt{e\_logg\_DR3} & float & log(cm s$^{-2}$) & $\sigma_{\rm{\logg}}$ from GALAH DR3 \\
    \texttt{flag\_sp\_DR3} & int & -- & Bitmask flag for \teff and \logg from GALAH DR3 \\
    \texttt{fe\_h\_DR3} & float & dex & \feh from GALAH DR3 \\ 
    \texttt{e\_fe\_h\_DR3} & float & dex & $\sigma_{\rm{\feh}}$ from GALAH DR3 \\
    \texttt{flag\_fe\_h\_DR3} & int & -- & Bitmask flag for metallicity from GALAH DR3 \\
    \texttt{snr\_DR3} & float & -- & S/N of the spectra from GALAH DR3 \\
    \hline
    \end{tabular}
    \label{tab:data}
\end{table*}

\begin{figure*}
    \centering
    \includegraphics[width=\textwidth]{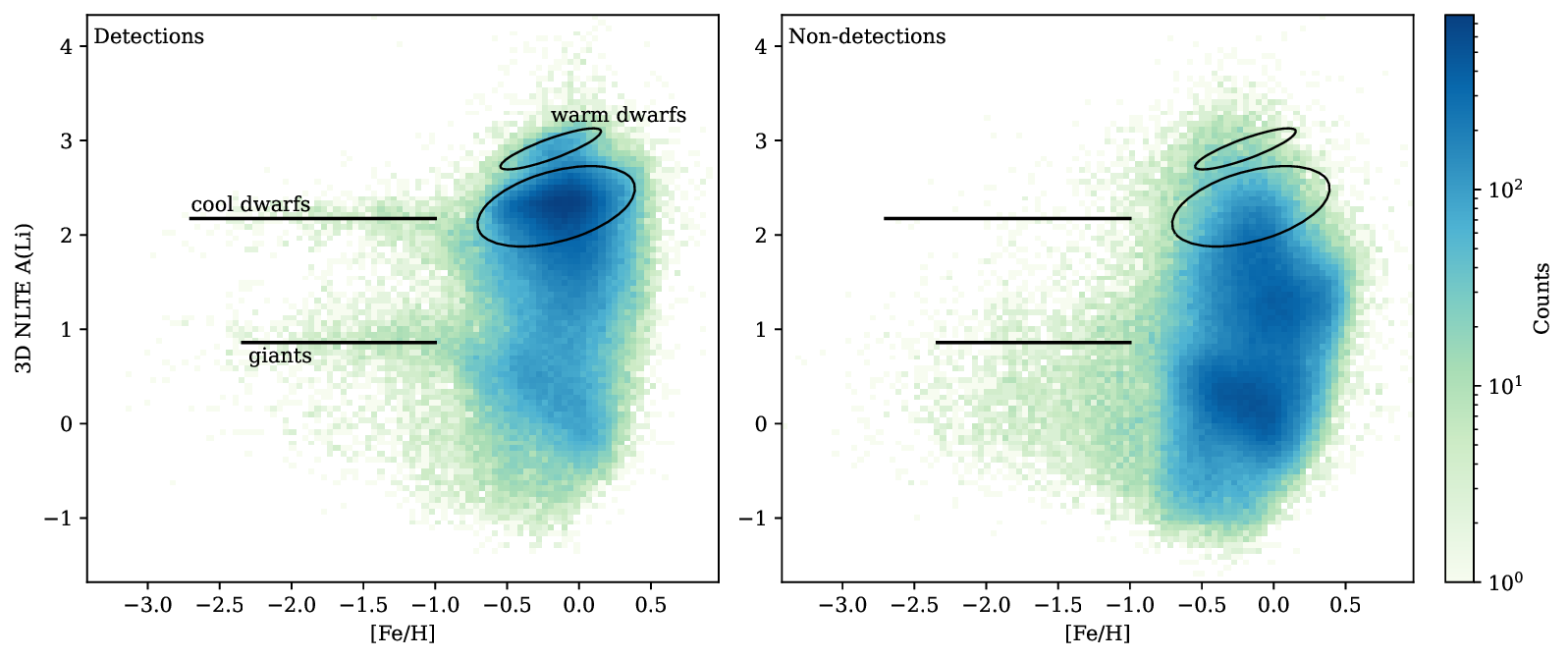}
    \caption{The distribution of 3D NLTE A(Li) detections (left) and upper limits (right) against [Fe/H]. There is an overdensity of warm and cool dwarfs, labelled as ovals. The cool plateau (also known as Spite plateau) and the metal-poor giant plateau are also labelled. There is also an overdensity of stars with non-detections in a similar location to the cool dwarfs with detections.
    A large number of metal-poor giants have upper limits at similar abundances to the metal-poor giant plateau.
    }
    \label{fig:feh}
\end{figure*}

Our derived 3D NLTE \ali over \feh are shown in Fig.~\ref{fig:feh} with detections on the left panel and non-detections with \ali reported as upper limits on the right panel. 
For stars with Li detections, we see an overdensity of warm and cool dwarfs for solar metallicity stars, labelled with ovals; and we see two plateaus: the metal-poor cool dwarf plateau, also known as the Spite plateau \citep{spite82, sbordone10}, and the metal-poor giant plateau \citep{mucciarelli12, mucciarelli14, mucciarelli22}; labelled with lines. These plateaus and overdensities are shown for illustrative purposes and not analysed further in this work. In the Spite plateau, detections and upper limits are well separated, with upper limits mostly below detections; this is in contrast to the metal-poor giant plateau, which has detections and upper limits at similar abundances. This indicates that for metal-poor dwarfs, we are distinguishing between stars with detected and non-detected Li successfully; whilst for metal-poor giants we are not. To distinguish between detections and non-detections for metal-poor giants, we need to lower the upper limits which can be achieved through higher S/N spectra. Given that detections and upper limits are very close in abundance for giants, we find that the current GALAH data cannot be used to constrain population synthesis modelling of Li abundance (similar to \citealt{chaname22}).

\begin{figure*}
    \centering
    \includegraphics[width=\textwidth]{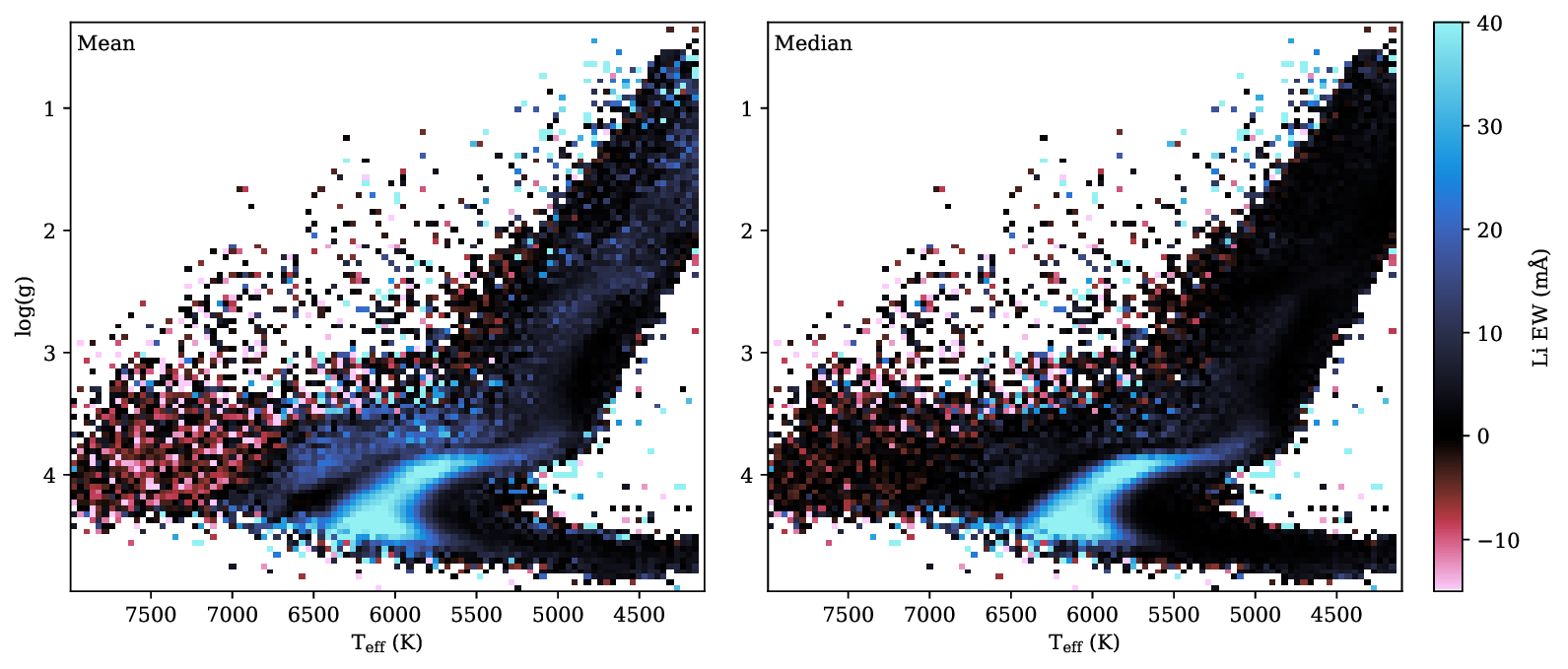}
    \caption{Mean EW (left) and median EW (right) binned in \teff and \logg for all stars in our sample, including non-detections. Both the mean and the median show stars with a higher Li EW in the MSTO at \teff$\approx$ 6000\,K and \logg$\approx$ 4.5, extending up the subgiant branch to \teff$\approx$ 5700\,K and \logg$\approx$ 4. Li-dip stars are hotter with lower Li EW at \teff$\approx$ 6500\,K and \logg$\approx$ 4.2, also extending up the subgiant branch. Li then increases in even hotter stars at \teff$\approx$ 7000\,K and \logg$\approx$ 4, however the Li EW does not reach as high as before the Li-dip. The mean can be lower than the median for rapid rotators at \teff$\approx$ 7500\,K; and the mean tends to be higher than the median for giant stars, likely due to the influence of Li-rich stars.}
    \label{fig:ave_ew}
\end{figure*}

We report the measured Li EW for all stars regardless of detection or non-detection. Fig.~\ref{fig:ave_ew} shows the variation of Li EW across \teff and \logg, with the left panel showing the mean and the right panel showing the median. Both panels show a region in the MSTO with high Li EW at \teff$\approx$ 6000\,K and \logg$\approx$ 4.5 extending up the subgiant branch to \teff$\approx$ 5700\,K and \logg$\approx$ 4. In addition, there is a region to the left with low Li EW at \teff$\approx$ 6500\,K and \logg$\approx$ 4.2; the well-known Li-dip. This region extends beyond the MSTO into the sub giant branch, and is likely the evolved Li-dip population. 
\edit{The transport of material out of the convective layer into the core of the star depletes Li and is stronger for cooler stars with a deeper convective zone. The Li-dip forms due to the moderation of this transport which inhibits Li depletion for stars cooler than the dip. The exact mechanism which inhibits this transport is unknown \citep{aguilera-gomez18}, with internal gravity waves \citep{charbonnel99, montalban00}, diffusion \citep{michaud86, richer93}, rotationally induced mixing \citep{pinsonneault90, deliyannis97}, and mass loss \citep{schramm90} being possible explanations.}
The Li-dip has been seen before in GALAH data \citep{gao20}, but this is the first time that a population of stars that have evolved onward from the Li-dip has been seen. Li is detected again for warm dwarfs hotter than the Li-dip, however the Li EW in stars hotter than the Li-dip is lower than for stars cooler than the dip. For stars hotter than the Li-dip, the mean is sometimes negative and lower than the median indicating that Li is not detected in a small number of stars, likely rapid rotators. The Li EW is weaker in giants compared to the MSTO, however, there are some scattered bins with a higher EW in the mean which are not present in the median. This feature is due to Li rich giants, which bring up the mean significantly. The red clump is visible in the mean Li EWs at \teff$\approx$ 4700\,K and \logg$\approx$ 2.5\,dex, indicating that there are a higher proportion of Li rich red clump stars compared to Li rich red giant branch stars. However, the median Li EW in the giant branch is low, indicating that neither population has significant Li retained. 

\begin{figure}
    \centering
    \includegraphics[width=0.48\textwidth]{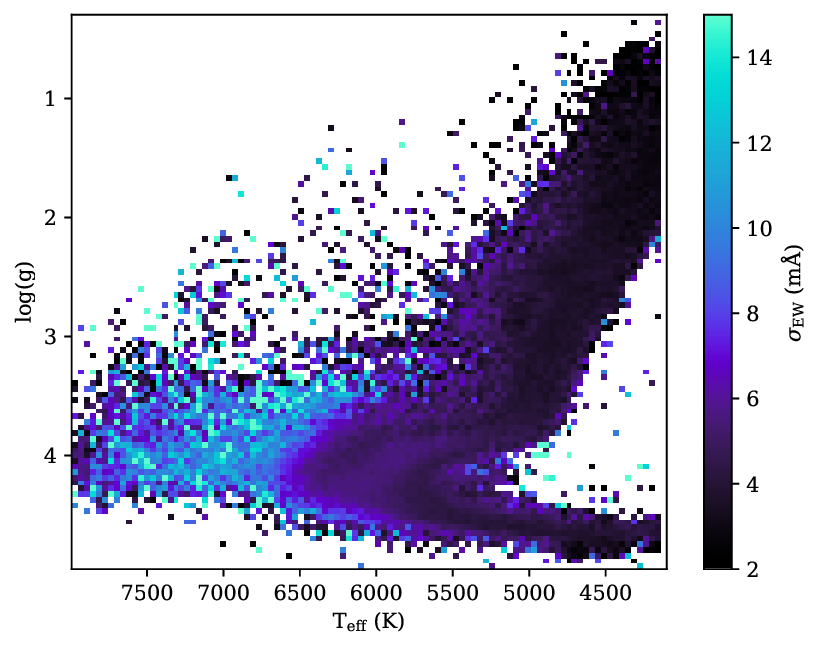}
    \caption{Mean error in Li EW, given by $(\sigma^{\rm l}_{\rm{EW}} + \sigma^{\rm u}_{\rm{EW}})/2$.
    In general, the error increases the hotter the star, with a jump in mean error for some stars hotter than \teff$\approx$ 6500\,K, mainly related to the rotational velocity. There is no strong correlation between \logg and the error in Li EW. 
    }
    \label{fig:ave_det_lim}
\end{figure}

We probe how the mean error in Li EW varies across \teff and \logg in Fig.~\ref{fig:ave_det_lim}. There is a sudden increase in Li EW error at \teff above 6500\,K, with some extremely high Li EW error bins scattered all throughout this high \teff MSTO region. This feature is due to rapid rotators, which have increased line blending due to high rotational broadening, thus making the line harder to measure as the depression from the line is lower. There is no strong trend in \logg, as the evolutionary state of the star influences the amount of line blending, we conclude that blending lines do not affect our Li EW error significantly. This is likely because we do not include model errors in our analysis. 

\begin{figure*}
    \centering
    \includegraphics[width=0.49\textwidth]{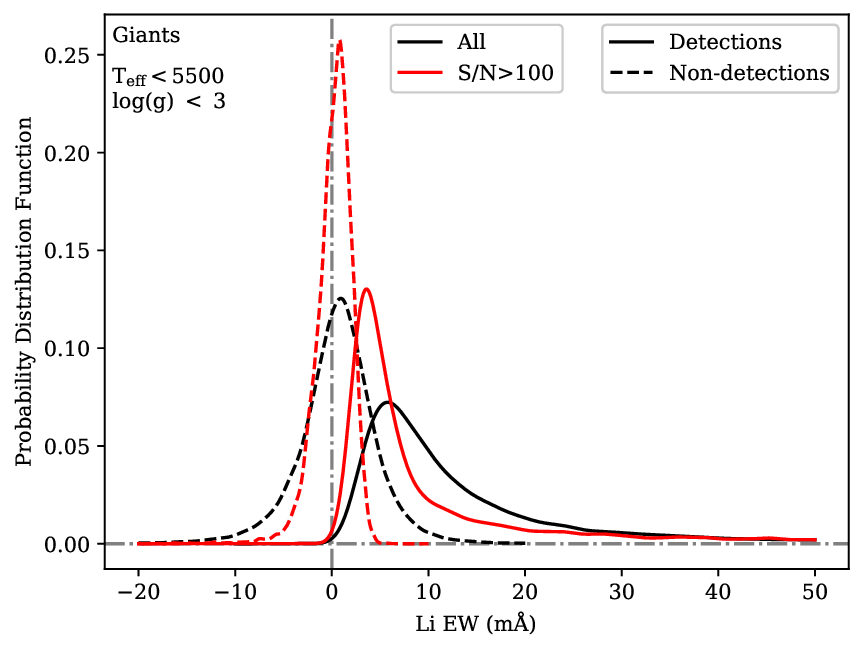}
    \includegraphics[width=0.49\textwidth]{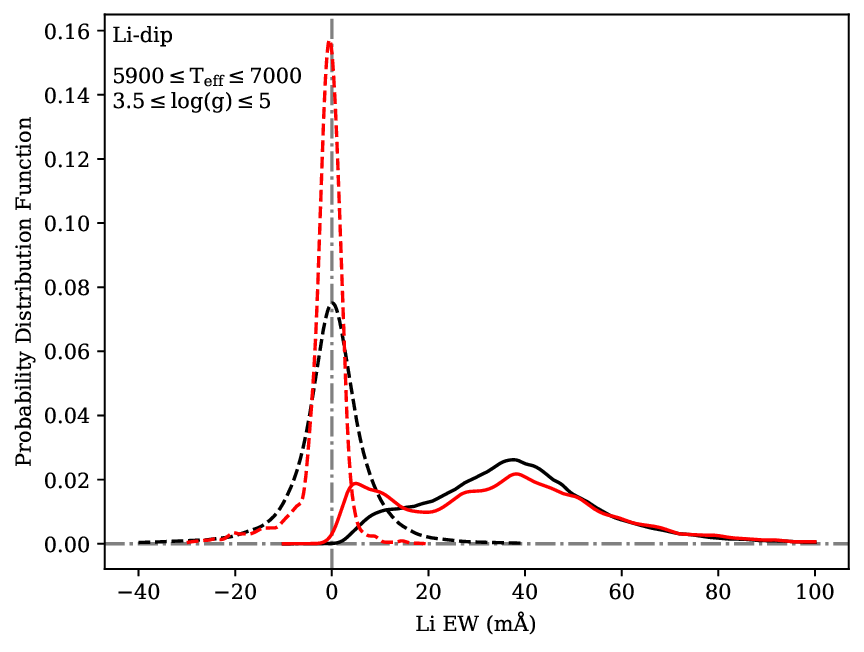}
    \caption{The probability density function of Li EW for giant stars (left) and Li-dip stars (right), showing both detections (solid) and non-detections (dashed). An additional S/N cut of 100 is made and shown in red. The grey dash dotted lines are at 0. Both giant and MSTO stars appear more well separated in detections and non-detections at higher S/N. However, giant stars detections are unimodal with a positive tail, whilst MSTO stars have bimodal detections.}
    \label{fig:sep}
\end{figure*}

We show how S/N affects detections and non-detections in Fig.~\ref{fig:sep}. The left panel shows giant stars, using \teff $<$ 5500\,K and \logg $>$ 3\,dex. An increase in S/N will increase the number of detections at lower EW and shift the overlap between detections and non-detections down in EW. Both these effects indicate that we are sensitive to weaker lines at higher S/N. The overlap in EW between detections and non-detections has a smaller range at higher S/N, indicating that we are better able to separate out detections and non-detections. \edit{However, the non-detections mode is slightly positive regardless of S/N, caused by some stars with a non-zero Li EW, indicating that we are not fully separating detections and non-detections for giant stars.}
There is a long \edit{detection} tail at high Li EW \edit{comprised of} Li rich giants. The smoothness of this tail indicates that there is no distinct cutoff between Li rich and Li normal giants, instead, giants exhibit the full range of Li abundances. The right panel shows stars within the Li-dip, taken as 5900 $\leq$ \teff $\leq$ 7000\,K and 3.5 $\leq$ \logg $\leq$ 5\,dex. Li-dip detections show a bimodality that is more evident at higher S/N because at higher S/N stars with less Li are detected. These high S/N detections are likely due to false positives which appear to outnumber the true positives significantly at these low EWs. False positives are when stars with no Li are measured with Li due to the noise in the spectra. We use a 2$\sigma$ detection limit, so this will occur for 2.5\% of stars with no Li in our data set. \edit{The non-detections mode is centered at zero, indicating that we are separating detections and non-detections for Li-dip stars. }

\section{Discussion}
\label{discussion}
We compare our 3D NLTE Li abundances to 1D NLTE Li abundances in GALAH DR3.
The flags used for this comparison are \texttt{flag\_sp\_DR3 == 0} and \texttt{flag\_fe\_h\_DR3 == 0}. Li EW that fall outside of the \breidablik grid are still accurate but these 3D NLTE Li abundances are extrapolated, so we use \texttt{flag\_ALi < 4} for Li EWs and \texttt{flag\_ALi < 2} for 3D NLTE Li abundances. For 1D NLTE abundances from GALAH DR3, we use \texttt{flag\_ALi\_DR3 < 2}. 

\begin{figure*}
    \centering
    \includegraphics[width=0.49\textwidth]{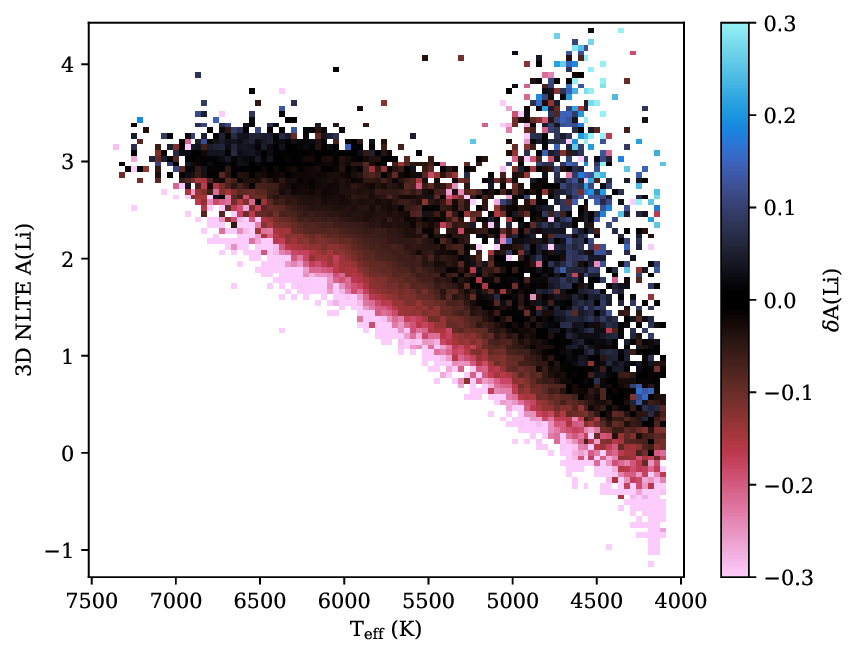}
    \includegraphics[width=0.49\textwidth]{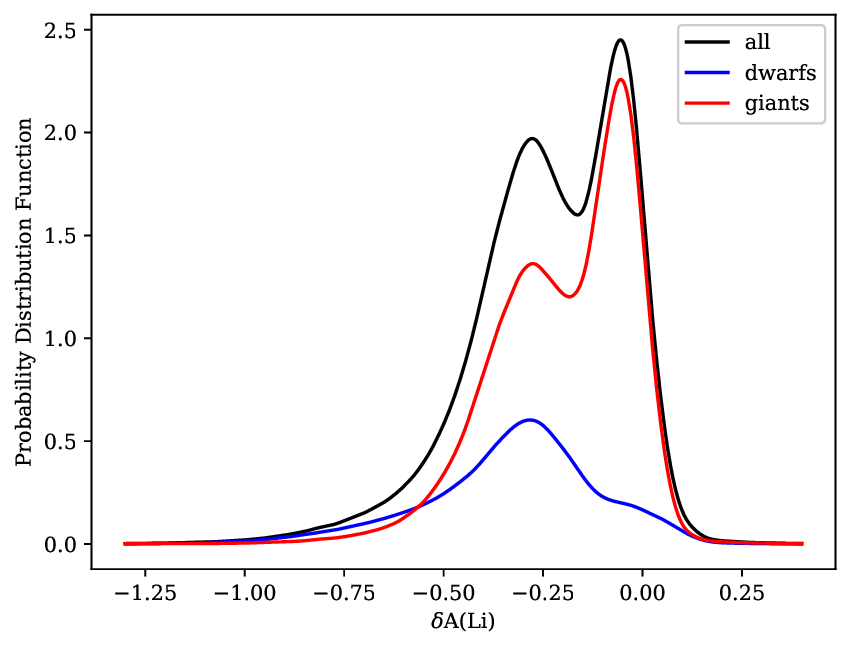}
    \caption{The difference between 3D NLTE \ali from this work and 1D NLTE \ali from GALAH DR3, defined as $\delta$\ali. The left panel shows how $\delta$\ali varies as a function of \teff and 3D NLTE \ali, truncated at 0.3 dex. The mean $\delta$\ali is negative with some positive $\delta$\ali at \teff$\approx$ 4500\,K. The increase in 3D NLTE \ali as \teff increases is due to the detection limit. Right panel shows $\delta$\ali as a histogram for giants compared to dwarfs. Here giants are stars with \logg$<$ 3.5\,dex and dwarfs are \logg$\geq$ 3.5\,dex. $\delta$\ali forms a bimodal distribution, with giants and dwarfs both contributing to the distribution centered at $\sim$-0.3 whilst giants mostly contribute to the distribution centered at $\sim$0.0. $\delta$\ali is not only due to 3D effects, but is also caused by a different analysis method.}
    \label{fig:diff}
\end{figure*}

The mean difference ($\delta$\ali) in \ali between this work and GALAH DR3 is shown in Fig.~\ref{fig:diff}. The differences here are not only due to 3D effects, but also due to a difference in analysis technique. GALAH DR3 fitted synthetic spectra with \textsc{SME}, which uses a linelist to set the blending line strengths with a scaled Solar composition\edit{, and only reported \ali where there were more than 5 unblended Li pixels}. In contrast, we have fitted spectra using a fine-tuned approach that emphasises the impact of blending lines and include errors in blends through sampling the posterior distribution. \edit{We report \ali regardless of number of blended Li pixels.} The left panel shows that in general our 3D NLTE abundances are lower than the GALAH DR3 abundances. This negative difference is primarily due to the difference in analysis technique and model spectra as the 1D NLTE to 3D NLTE correction for \ali$\approx 2$ dex is positive for effectively the entire HR diagram (derived using data from \citealt{breidablik}). The Li abundance increases as \teff increases due to the detection limit, where a higher \ali is required to form a line of the same Li EW at higher \teff. Our abundances can be higher than the GALAH DR3 abundances, this mainly occurs for giant stars at \teff$\approx$ 4500\,K, for a range of Li abundances. Furthermore, our positive differences do show a slight trend in Li abundance, where the difference is more positive for higher Li abundance. These stars with a positive difference tend to be giant stars with a line depression of $\sim$20\%, resulting in a positive correction. 
The right panel shows the distribution of $\delta$\ali for dwarfs and giants, where we take \logg$<$ 3.5\,dex as giants and \logg$\geq$ 3.5\,dex as dwarfs. On average, our abundances differ from GALAH DR3 by $-0.23$\,dex. The all stars distribution shows a bimodality in difference: at $\sim$-0.3, due to both giants and dwarfs; and at $\sim$0.0, primarily due to giants. Similar to the left panel, this shows that on average our differences are negative, however, we do have a small number of positive differences primarily for giant stars. 

\begin{figure}
    \centering
    \includegraphics[width=0.48\textwidth]{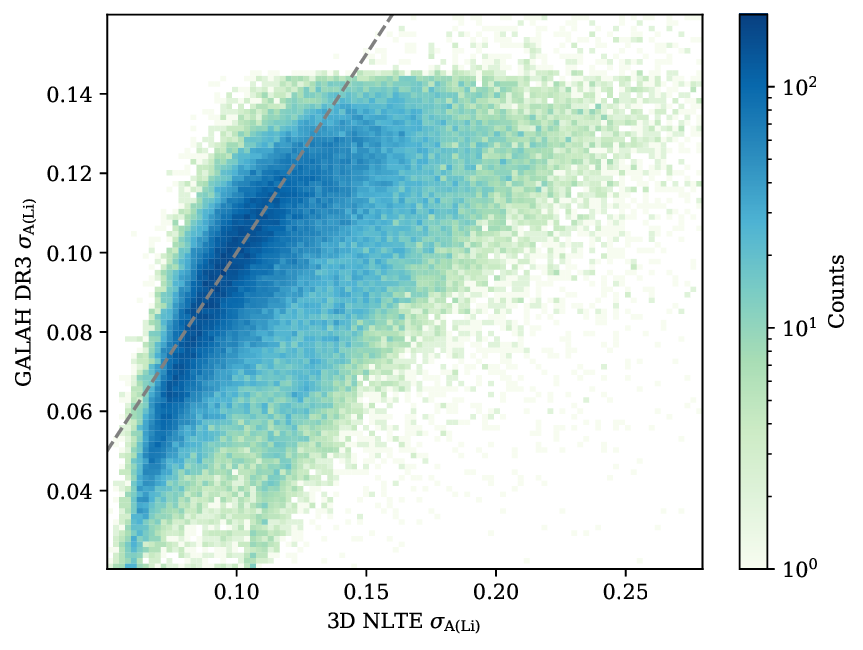}
    \caption{Comparison of error in \ali derived in this work to GALAH DR3. The errors are comparable, with the median at 0.1\,dex for both data sets. There is a slight curvature in the comparison, where for low and high errors in GALAH DR3, we derive higher errors. There is a secondary track where we derive higher errors than GALAH DR3. In addition, there is a sharp cutoff in GALAH DR3 errors that is not present in our errors. The gray dashed line show where the two errors are equal.}
    \label{fig:comp_err}
\end{figure}

We compare our errors in \ali against GALAH DR3 in Fig.~\ref{fig:comp_err} for stars with detected \ali. To derive the error in \ali, 3D NLTE $\sigma_{\rm{A(Li)}}$, we take the mean error due to EW, summed in quadrature with the error in \teff. The median error in this work is similar to GALAH DR3, at $\sigma_{\rm{A(Li)}} \approx 0.1$\,dex. However, there is a curvature in the errors where we derive higher errors for stars with both low and high GALAH DR3 errors. This curvature at low GALAH DR3 errors is due to our errors having a Li EW dependence whilst GALAH DR3 errors do not. There is a sharp cutoff in GALAH DR3 errors at $\sigma_{\rm{A(Li)}} \approx 0.15$\,dex because stars with larger error are non-detections. This cutoff is causing the curvature at high GALAH DR3 errors. A similar cutoff is not present for our errors as we do not incorporate an error cutoff for detections. The secondary track where our errors are higher than GALAH DR3 errors are caused by large $\sigma_{\rm{T}_{\rm{eff}}}$ values, driving up our errors. 

\begin{figure*}
    \centering
    \includegraphics[width=0.49\textwidth]{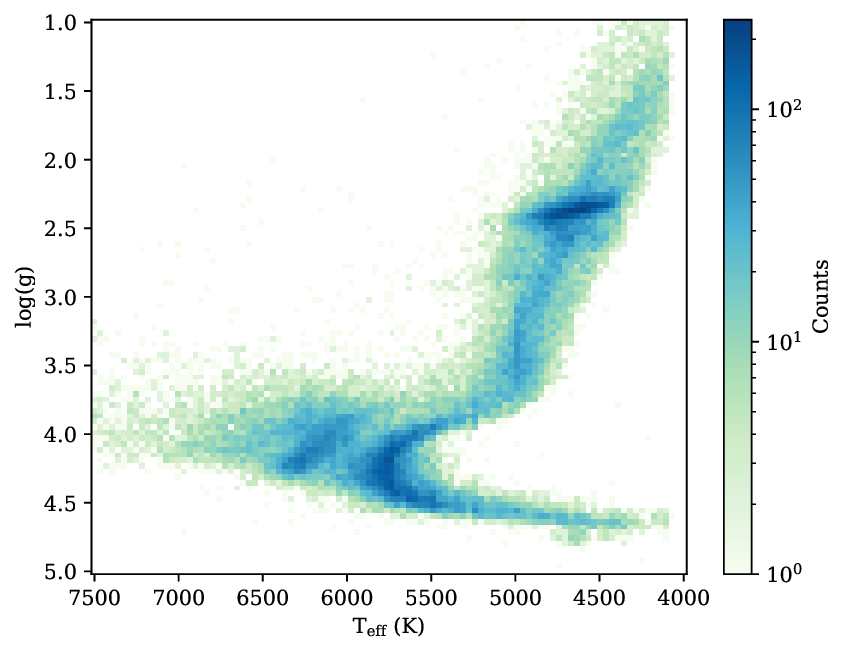}
    \includegraphics[width=0.49\textwidth]{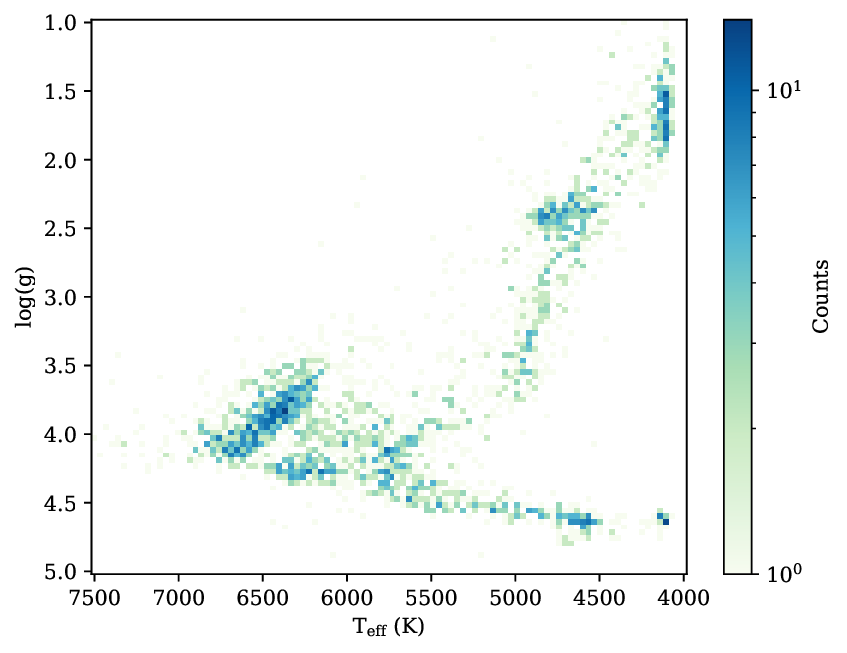}
    \caption{Stars that have detections in this work but not in DR3 (left) compared to stars that have detections in DR3 but not this work (right). This work increases detections for all stellar parameters compared to DR3, however, we increase detections most in RC stars, and MSTO stars that are cooler or hotter than the Li dip. Compared to DR3, we lose detections mostly in the hot MSTO stars, where the rapid rotators are.}
    \label{fig:comp}
\end{figure*}

Fig.~\ref{fig:comp} compares detections between our data set and GALAH DR3. This work detects Li in 50048 more stars compared to DR3 over all stellar parameters, shown in the left panel. The increase in detections is in part due to differences in the detection criteria used and the difference in model spectra. The density of our detections for dwarf stars are similar to Fig.~\ref{fig:ave_ew}, indicating that we detect more stars with Li where the mean Li EW is higher. This is not true for giants, where we detect more stars with Li in the red clump than GALAH DR3, however these stars do not have a higher mean Li EW. Detections in GALAH DR3 but not this work traces a similar shape, as shown in the right panel. However, GALAH DR3 detects more stars with Li in hot MSTO stars, likely the rapid rotators where our analysis has a stricter detection cutoff compared to GALAH DR3. In addition to these rapid rotators, GALAH DR3 also detects more stars with Li at \teff$\approx$ 4100\,K\edit{, likely due to identified overdensities in the GALAH DR3 parameter space \citep{DR3}.}

\section{Conclusion}
\label{conclusion}
In this paper we publish 3D NLTE Li abundances and equivalent widths for \stars stars observed in GALAH DR3. For these stars, we measure Li abundances for \detection stars and provide upper limits for \upperlim stars. 
We fit the Li region of the observed spectra using model spectra produced by multiplying together a 3D NLTE Li profile and Gaussian profiles for other lines. We provide errors that consider blending lines by sampling the posterior for all our stars using these model spectra. We validate this method on synthetic spectra and by comparison to 26 stars in the Hypatia catalogue. 
We identify the known Li phenomena, including the Spite plateau formed by cool metal-poor dwarfs, the metal-poor giant plateau, an overdensity of warm and cool dwarfs at Solar metallicity, and the Li-dip. 
Using our Li EWs, we find stars that evolved from the Li-dip on the main sequence turn-off into the subgiant branch.
We show that the error in Li EW increases sharply for rapid rotators and the error in Li abundance has a \teff dependence. 
Non-detections are defined through our Li EW errors, which are correlated with S/N. We show that higher S/N better separates detections and non-detections. 
Comparing our abundances with 1D NLTE GALAH DR3 Li abundances, on average we differ by 0.23\,dex in measured abundance, with majority of our abundances lower than GALAH DR3. We find the detection limit, which is strongly influenced by EW, increases as \teff increases. 
In addition, we find that our errors are similar to GALAH DR3, with median error of $\sigma_{\rm{A(Li)}} \approx 0.1$\,dex for both. 
We detect Li in more stars than GALAH DR3, most of these stars fall in regions with higher mean Li EW; whilst GALAH DR3 detects Li in more rapid rotators compared to this work.

In future work, we plan on implementing this method for the upcoming GALAH data release 4 and using it to investigate various Li phenomena identified during our analysis. \edit{We note that the methodology presented in this paper is applicable with minimal modification to future surveys e.g. WEAVE \citep{dalton16} and 4MOST \citep{dejong19}.}

\section{Data Availability}
The data produced as part of this work, described in Section~\ref{results}, can be downloaded as a Value Added Catalogue from the GALAH DR3 website, \url{https://www.galah-survey.org/dr3/the_catalogues/#3d-nlte-li-abundances}.
The sampled posteriors can be provided upon request. 
The code described in Section~\ref{analysis} can be found at \url{https://github.com/ellawang44/galah-li}.

\section*{Acknowledgements}
\edit{We thank the anonymous referee for their many suggestions improving the quality and content of this manuscript.}
This work was supported by computational resources provided by the Australian Government through the National Computational Infrastructure (NCI) under the National Computational Merit Allocation Scheme and the ANU Merit Allocation Scheme (project y89) and HPC-AI Talent Program Scholarship (project hl99).
This work was supported by the Australian Research Council Centre of Excellence for All Sky Astrophysics in 3 Dimensions (ASTRO 3D), through project number CE170100013. SLM acknowledges support from the Australian Research Council through Discovery Project grant DP220102254, and from the UNSW Scientia Fellowship Program.
KL acknowledges funds from the European Research Council (ERC) under the European Union’s Horizon 2020 research and innovation programme (Grant agreement No. 852977) and funds from the Knut and Alice Wallenberg Foundation.
We acknowledge the traditional owners of the land on which GALAH data was taken, the Gamilaraay people, and pay our respects to elders past, present, and emerging.




\bibliographystyle{mnras}
\bibliography{ref}



\appendix

\section{\breidablik}
\label{app:b}
We update the line profile interpolation routine in \breidablik to use radial basis functions (RBF) instead of Kriging, taking the implementation of RBF from \citet{bertran22}. The methodology used for comparison is the same as presented in \citet{breidablik}. 

To make the interpolation easier, we transform the flux using
\begin{equation}
    f_{\rm{t}} = \log_{10}(1 - f + s)
\end{equation}
where $s$ is a small positive constant that is treated as a hyperparmeter in the interpolation methods. For RBF, we find $s=10^{-5}$ using 5-fold cross-validation. 

We compare the performance of RBF against Kriging, taking the Kriging results from \citet{breidablik}. The leave-one-out cross-validation errors, interpolation, extrapolation, training, and execution time of the code is shown in Table~\ref{tab:errors}. Both the mean absolute deviation (MAD), which is less affected by outliers, and the root mean squared (RMS), which is more affected by outliers, are reported. The 3D leave-one-out cross validation errors are likely overestimates as the step sizes in this test were 500\,K \teff, 0.5\,dex \logg, and 0.5\,dex \feh, whilst in practice, interpolating on the full 3D grid will use step sizes closer to 250\,K \teff, 0.25\,dex \logg, and 0.25\,dex \feh. To account for this, we also measure leave-one-out cross validation errors for a 1D grid, as doubling the size of our 3D grid to do these tests is computationally infeasible. These 1D tests include 1000 interpolation points, where the data points are fully bounded by existing grid points, and 1000 extrapolation points, where the data points are not fully bounded by the existing grid. These results show that RBF is comparable to Kriging for the 3D grid and 1D interpolation, slightly better for 1D extrapolation, and is faster to train and execute.

\begin{table*}
\caption{Comparison of Kriging and RBF interpolation errors when measuring A(Li) from the 670.8\,nm line profile. The MAD and RMS error statistics for A(Li) are reported using leave-one-out cross-validation on the 3D grid, and interpolation and extrapolation tests on a 1D grid. The time required to train the model and the time required to produce the three Li lines (610.4\,nm, 670.8\,nm, 812.6\,nm) with arbitrary values of \teff, \logg, \feh, and A(Li) is also recorded.}
\begin{tabular}{c|cccccccc}
& \multicolumn{2}{c}{3D leave-one-out} & \multicolumn{2}{c}{1D interpolation} & \multicolumn{2}{c}{1D extrapolation} \\
Method & MAD & RMS & MAD & RMS & MAD & RMS & training time (s) & execution time (s) \\ 
\hline
Kriging & 0.014 & 0.022 & 0.020 & 0.038 & 0.019 & 0.143 & 235 & 5.62 \\
RBF & 0.012 & 0.018 & 0.022 & 0.039 & 0.016 & 0.062 & 56.3 & 0.0273 \\
\hline
\end{tabular}
\label{tab:errors}
\end{table*}

\begin{figure*}
    \centering
    \includegraphics[width=0.49\textwidth]{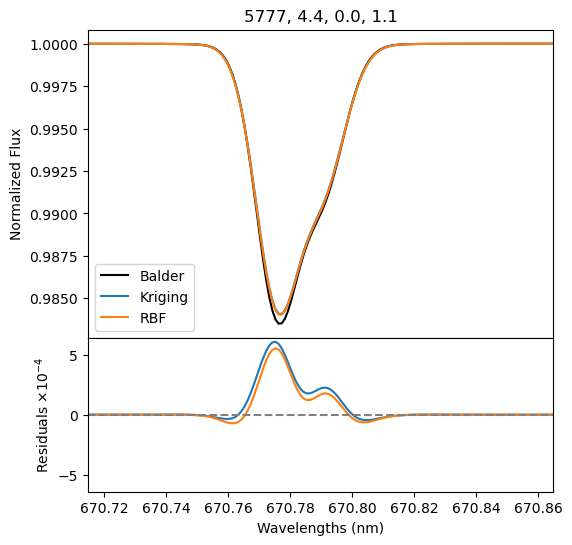}
    \includegraphics[width=0.49\textwidth]{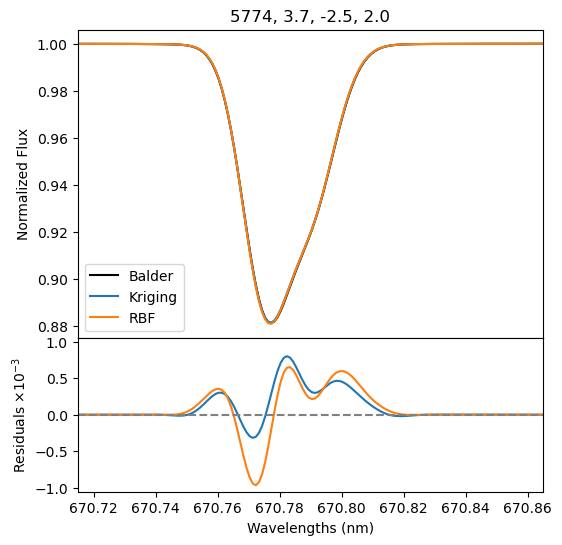}
    \includegraphics[width=0.49\textwidth]{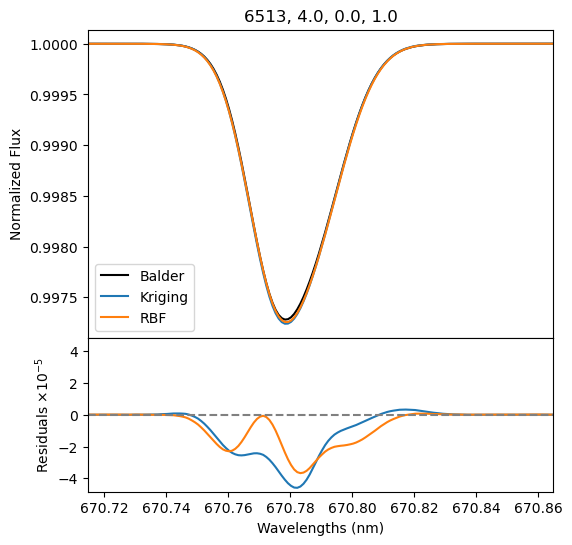}
    \includegraphics[width=0.49\textwidth]{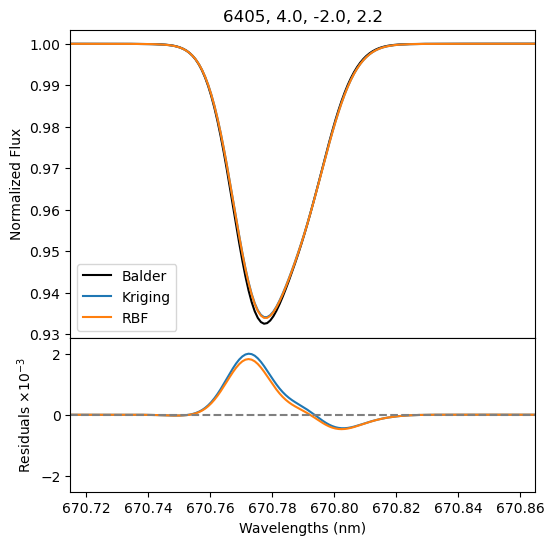}
    \caption{Comparison between RBF and Kriging on verification models (Balder) that were not trained on.These models were tailored to specific stars, indicated by name and stellar parameters: \teff, \logg, \feh, and A(Li). Note that the residuals between prediction and verification model have been magnified by a different amount in each plot.}
    \label{fig:ver}
\end{figure*}

To verify the final interpolation model and to demonstrate RBF interpolated line profiles and the usual expected errors, we measured the abundances of 4 benchmark stars. Again, we compare these results to Kriging from \citet{breidablik}. The measured abundances are are shown in Table~\ref{tab:ver_prof}, and the corresponding interpolated line profiles for the simulated A(Li) are shown in Fig.~\ref{fig:ver}. For these benchmark stars, RBF outperforms Kriging slightly. 

\begin{table*}
\caption{A comparison between Kriging and RBF interpolation on benchmark stars. Direct synthesis in the first row is the reference A(Li), subsequent rows is the error relative to the reference. The verification models were tailored to specific stars, indicated by name, whose stellar parameters are given in Fig.~\ref{fig:ver}.}
\begin{tabular}{c|cccc}
\diagbox{Method}{Model} & Sun & HD 140283 & Procyon & HD 84937 \\
\hline
Direct synthesis & 1.100 & 2.000 & 1.000 & 2.200 \\
\hline
Kriging & 0.012 & 0.002 & -0.006 & 0.009 \\
RBF & 0.010 & 0.000 & 0.007 & -0.004 \\
\hline
\end{tabular}
\label{tab:ver_prof}
\end{table*}

\section{Additional Fits}
\label{app:fits}

As examples, we provide addition fits to: a warm solar metallicity dwarf with low S/N, a poorly constrained fit, and a saturated fit. 

\begin{figure*}
    \centering
    \includegraphics[width=\textwidth]{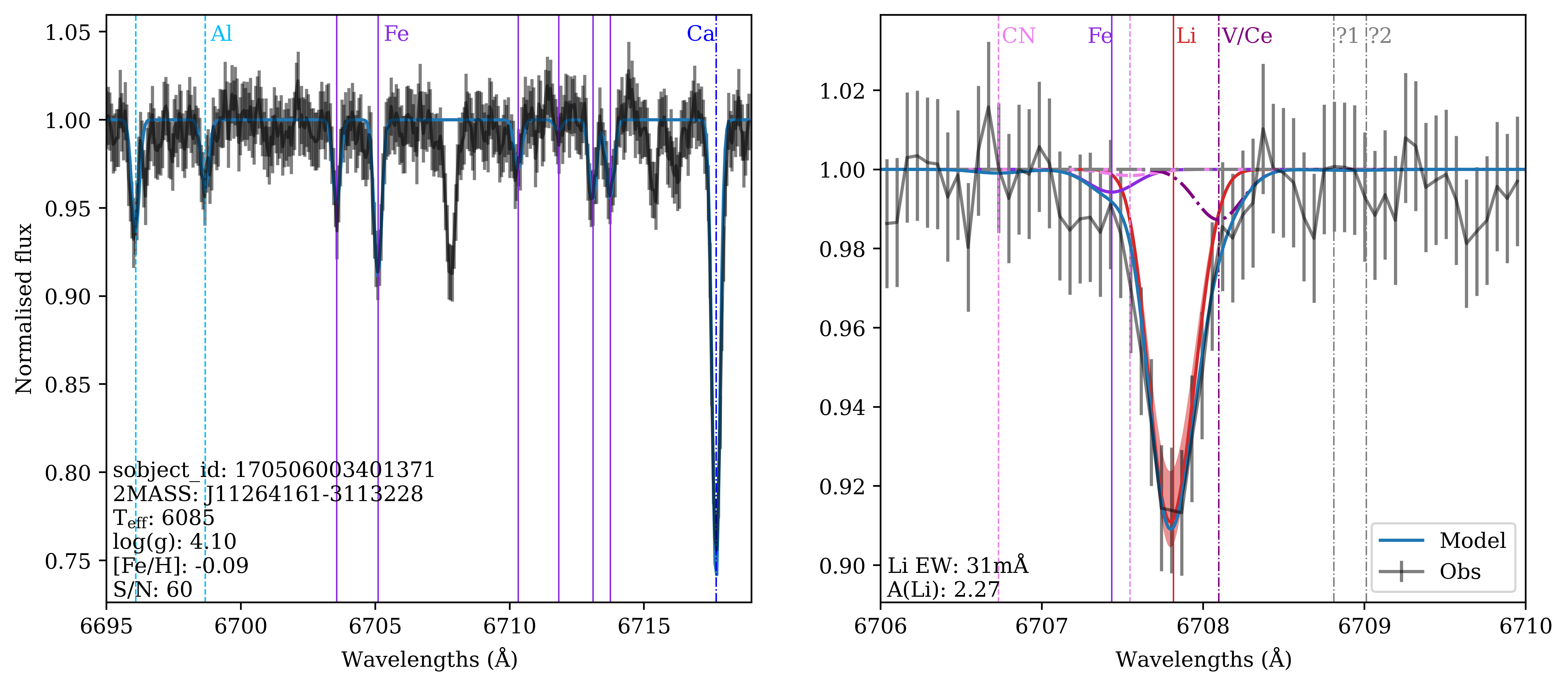}
    \caption{Example fit of a warm solar metallicity dwarf with low S/N, where the lithium line dominates over the blending lines. Labels follow Fig.~\ref{fig:fit}.}
\end{figure*}

\begin{figure*}
    \centering
    \includegraphics[width=\textwidth]{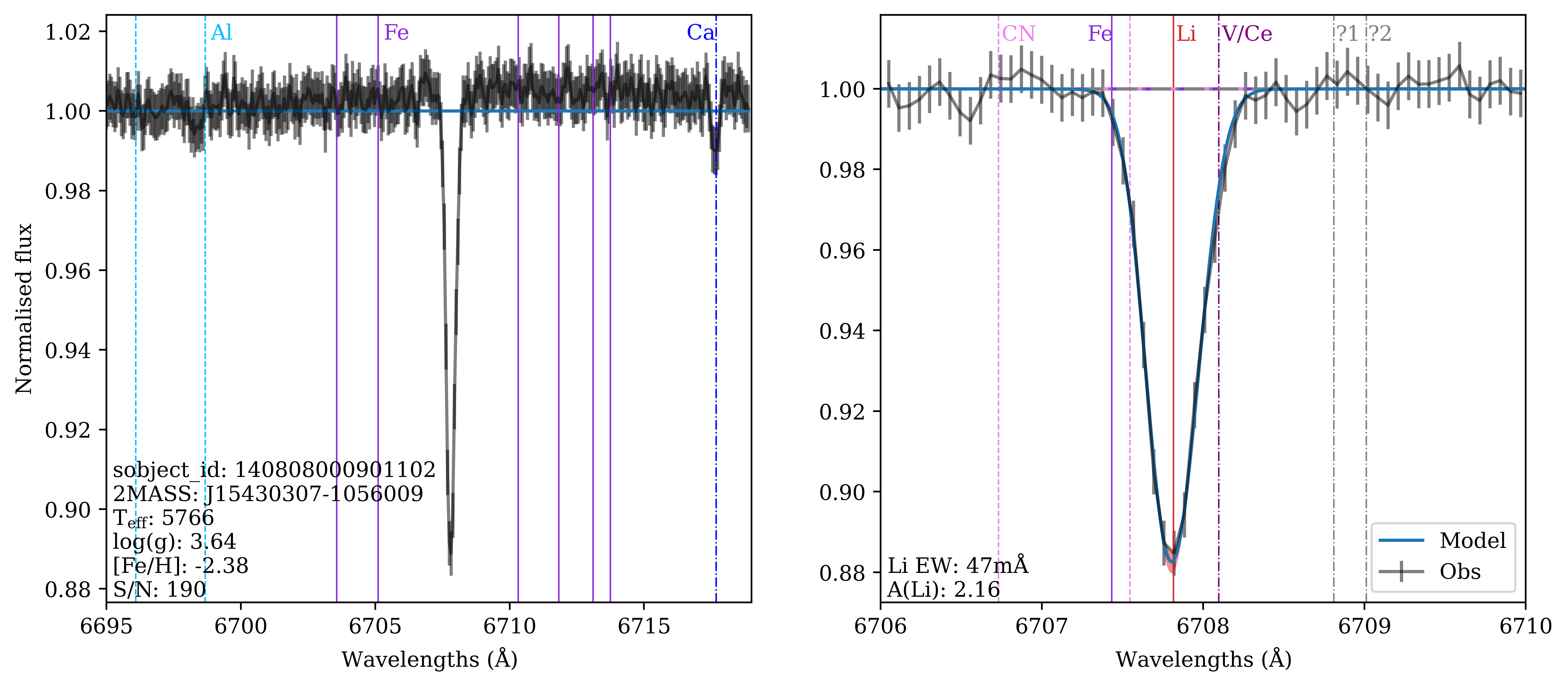}
    \caption{Example fit of a poorly constrained star, where there are less than 3 lines above the noise in the broad region, labeled like Fig.~\ref{fig:fit}.}
\end{figure*}

\begin{figure*}
    \centering
    \includegraphics[width=\textwidth]{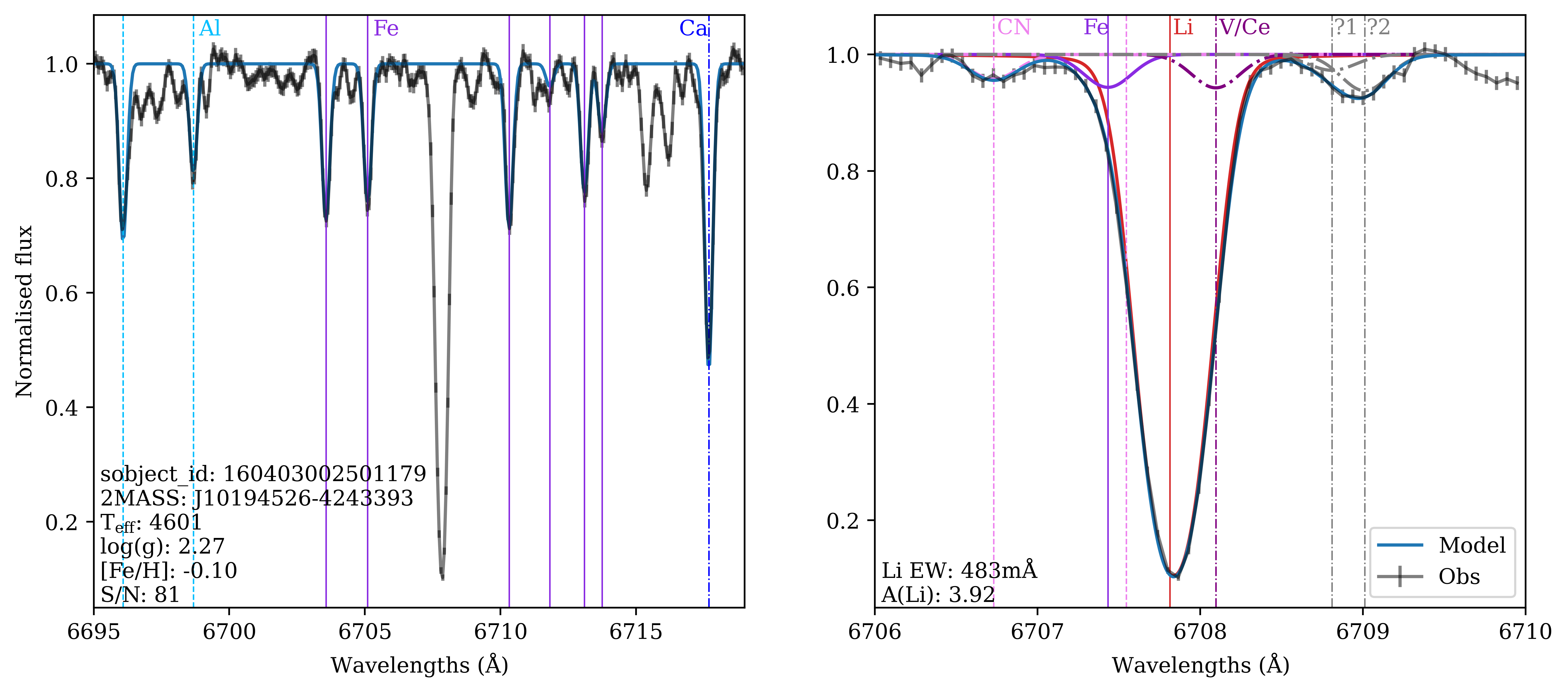}
    \caption{Example of a strongly saturated star, indicating an excellent fit with the synthetic spectrum. Labels follow Fig.~\ref{fig:fit}.} 
\end{figure*}

\section{Posterior error}
\label{app:err}

We do not fully capture the sampled posterior for 170 stars due to restrictions from the prior, shown in the right panel of Fig.~\ref{fig:err}. \edit{T}he highest posterior density Bayesian credible interval does not exist \edit{for these sampled posteriors}. \edit{In these cases}, we report 34\% below or above the MAP as the lower or upper error, depending on which one is defined. The corresponding undefined lower or upper error is then set to NaN.
\begin{figure}
    \centering
    \includegraphics[width=0.48\textwidth]{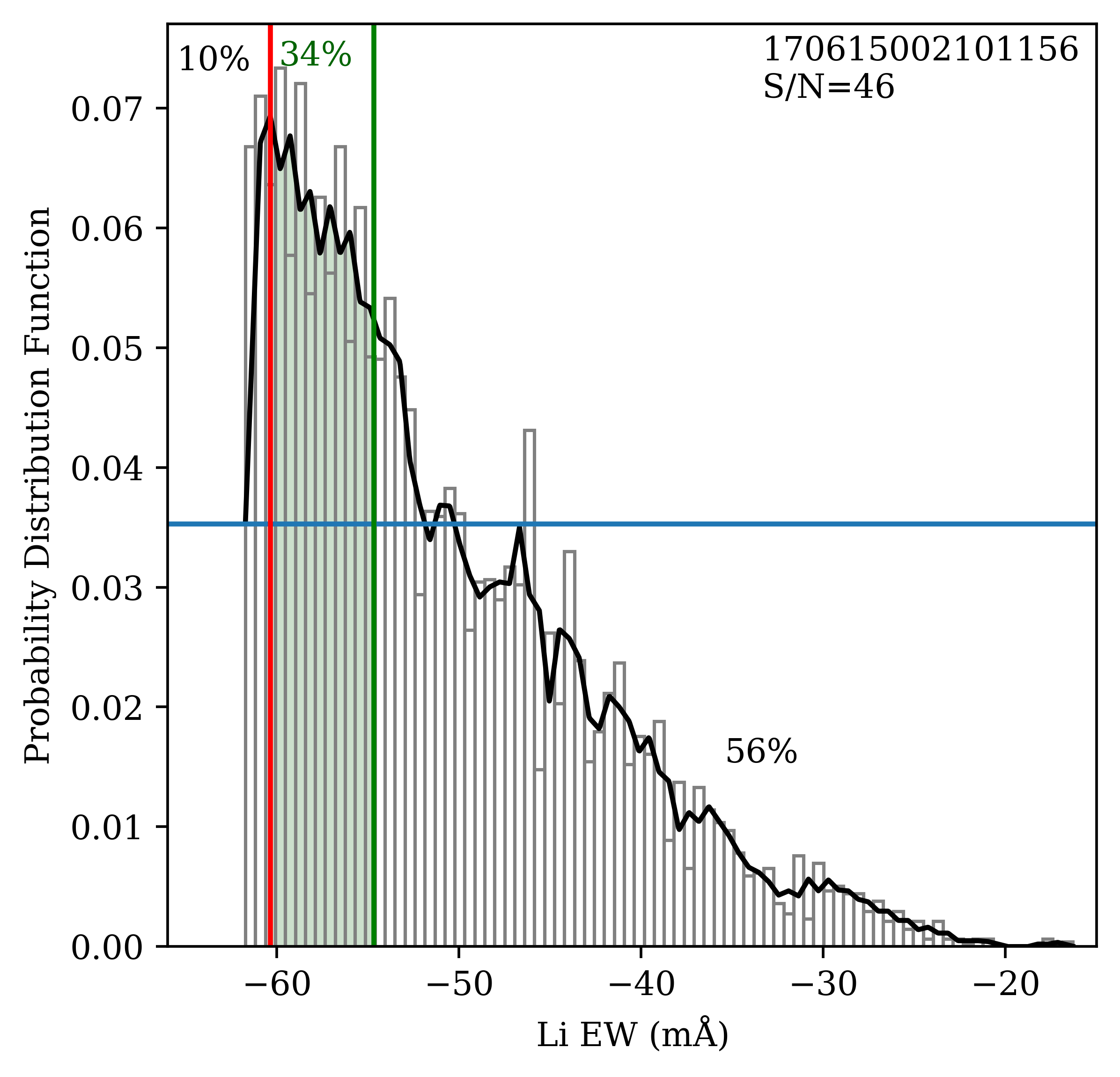}
    \caption{Example sampled posterior showing how we derive errors for a posterior with one side truncated. The sampled posterior is still binned into a histogram (black bars) and smoothed (solid black line), with the MAP (red) adopted as the measured Li EW. However, the horizontal line (blue) encompassing 68\% of the pdf only intercepts the pdf once, so the upper error is reported as 34\% above the MAP (green).}
    \label{fig:err_special}
\end{figure}

\section{Synthetic Spectrum Validation}
\label{app:synth}
\begin{figure*}
    \centering
    \includegraphics[width=\textwidth]{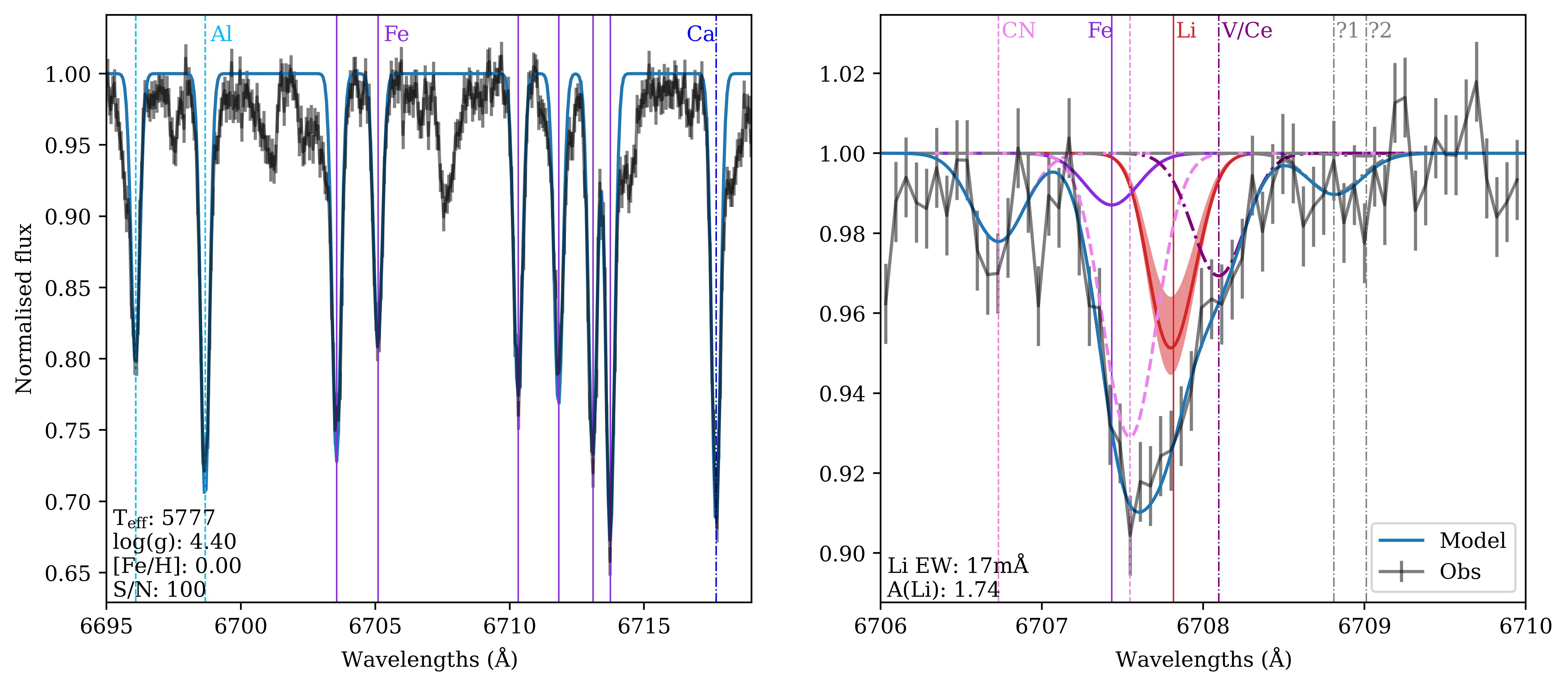}
    \caption{Example fit to a synthetic spectrum with added noise. This spectrum is intended to be similar to a heavily blended Solar spectrum. Labels follow Fig.~\ref{fig:fit}.}
    \label{fig:synth}
\end{figure*}

We validate our method on synthetic spectrum. Fig.~\ref{fig:synth} shows the synthetic spectrum that we test on and our corresponding best fit. 

The input parameters and the fitted results are shown in Table~\ref{tab:synth}.

\begin{table}
    \centering
    \caption{The true and measured parameters for validation on a synthetic spectrum. The measured parameters are close to the true parameters, with the broad region measured parameters usually larger than the true parameters.}
    \begin{tabular}{cccc}
    \hline 
    Parameter & Units & True & Measured \\
    \hline
    \multicolumn{4}{c}{Broad region} \\
    \hline
    Al I 6696.085 & m\AA & 75 & 83 \\   
    Al I 6698.673 & m\AA & 108 & 122 \\ 
    Fe I 6703.565 & m\AA & 82 & 112 \\ 
    Fe I 6705.101 & m\AA & 77 & 79 \\
    Fe I 6710.317 & m\AA & 87 & 93 \\
    Fe I 6711.819 & m\AA & 74 & 96 \\ 
    Fe I 6713.095 & m\AA & 104 & 110 \\
    Fe I 6713.742 & m\AA & 110 & 134 \\ 
    Ca I 6717.681 & m\AA & 119 & 130 \\
    FWHM & \kms & 17.13 & 17.22 \\
    \vr & \kms & 3.52 & 3.38 \\
    \hline
    \multicolumn{4}{c}{Narrow region} \\
    \hline
    CN 6706.730 & m\AA & 7.2 & 9.1 \\
    Fe I 6707.433 & m\AA & 13 & 5.3 \\
    CN 6707.545 & m\AA & 19 & 29 \\
    Li I 6707.814 & m\AA & 20 & 17 \\
    V/Ce 6708.096 & m\AA & 8.4 & 13 \\
    ?1 6708.810 & m\AA & 1.3 & 4.1 \\
    ?2 6709.011 & m\AA & 1.6 & 0.3 \\
    FWHM$_{\rm{Li}}$ & \kms & 13.68 & 10.09 \\
    \hline
    \end{tabular}
    \label{tab:synth}
\end{table}


\bsp	
\label{lastpage}
\end{document}